\documentclass[a4paper,journal]{IEEEtran}
\usepackage{amsmath,amsfonts,amssymb}
\usepackage{algorithmic}
\usepackage{array}
\usepackage{textcomp}
\usepackage{stfloats}
\usepackage{url}
\usepackage{verbatim}
\usepackage{graphicx}
\usepackage{balance}
\usepackage{paralist}
\usepackage{booktabs}

\usepackage{tikz}
\usepackage{pgf}
\usepackage{pgfplots}
\usepackage{tikzscale}
\usepackage{calc}
\usetikzlibrary{calc}
\usepackage{tikz-dimline}

\usepackage{caption}
\usepackage{subcaption}
\usepackage{cite}

\newcommand{\define}[1]{\emph{#1}}

\newcommand{\vect}[1]{\ensuremath{\mathbf #1}}

\newcommand{\set}[1]{\ensuremath{\left\{ #1 \right\}}}
\newcommand{\setof}[2]{\ensuremath{\left\{ #1 : #2 \right\}}}

\begin{document}

\title{DCP and VarDis: An Ad-Hoc Protocol Stack for Dynamic Swarms and
  Formations of Drones\\Extended Version}

\author{
  \IEEEauthorblockN{Samuel Pell, Andreas Willig}\\
  \IEEEauthorblockA{\textit{Department of Computer Science} \\
    \textit{University of Canterbury}\\
    Christchurch, New Zealand \\
    sam.pell@canterbury.ac.nz, andreas.willig@canterbury.ac.nz}%
}

\maketitle

\begin{abstract}
  Recently, swarms or formations of drones have received increased
  interest both in the literature and in applications. To dynamically
  adapt to their operating environment, swarm members need to
  communicate wirelessly for control and coordination tasks. One
  fundamental communication pattern required for basic safety
  purposes, such as collision avoidance, is beaconing, where drones
  frequently transmit information about their position, speed,
  heading, and other operational data to a local neighbourhood, using
  a local broadcast service. In this paper, we propose and analyse a
  protocol stack which allows to use the recurring-beaconing primitive
  for additional purposes. In particular, we propose the VarDis
  (Variable Dissemination) protocol, which creates the abstraction of
  variables to which all members of a drone swarm have (read) access,
  and which can naturally be used for centralized control of a swarm,
  amongst other applications. We describe the involved protocols and
  provide a mainly simulation-based performance analysis of VarDis.
\end{abstract}



\bstctlcite{IEEEexample:BSTcontrol} 

\section{Introduction}
\label{sec:introduction}

Unmanned aerial vehicles (UAVs) or drones are nowadays used in a wide
range of commercial or scientific applications, including
environmental monitoring and remote sensing
\cite{Gevaert:Suomalainen:Tang:Kooistra:15,Saffre:Hildmann:Karvonen:Lind:22},
infrastructure monitoring, search and rescue
\cite{Barry:Willig:Woodward:22,Careem:Gomez:Saha:Dutta:21},
surveillance, telecommunications and public safety communications
\cite{Geraci:GarciaRodriguez:Azari:Lozano:EtAl:22}, agriculture
\cite{Caruso:Chessa:Escolar:Barba:Lopez:21,Li:Long:Wu:Hu:Wei:Zhang:Chai:Xie:Mei:22}
and forestry \cite{LewisHartley:Henderson:Jackson:22}, delivery of
small goods in urban areas
\cite{ValenciaArias:RodriguezCorrea:PatinoVanegas:BenjumeaArias:DeLaCruzVargas:MorenoLopez:22},
and many others. Many of these applications focus on individual
drones, but there is also increasing interest in applications and
communications / networking support for swarms or formations of
closely collaborating drones
\cite{Baltaci:Dinc:Ozger:Alabbasi:Cavdar:Schupke:21,Chandrasekharan:Gomez:AlHourani:Kandeepan:Rasheed:Goratti:Reynaud:Grace:Bucaille:Wirth:Allsopp:16,Ghamry:Kamel:Zhang:16,GuillenPerez:Cano:18,Gao:Li:Cai:22,Saffre:Hildmann:Karvonen:Lind:22,Zhang:Chermprayong:Xiao:Tzoumanikas:EtAl:22}. Drone
swarms naturally allow for parallelization of a task. For example, in
a search-and-rescue operation, to speed up the overall search process,
the search area can be partitioned and each drone is assigned to one
such partition. Beyond this, using multiple drones allows observation
of the same phenomenon from different angles or using different
sensors at the same time, or to form flying antenna arrays supporting
localization of transmitters
\cite{Pell:Willig:22,Werner:Wang:Hakkarainen:Gulati:Patron:Pfeil:Dandekar:Cabric:Valkama:15}
or telecommunications.

Depending on the application, drone swarms can require significant
amounts of coordination and data exchange between drones. A very basic
(and application-independent) requirement is that drones are able to
keep track of the position, speed and heading of nearby drones to
avoid collisions, particularly in the presence of disturbances such as
wind gusts. The knowledge of position / speed / heading of neighbours
is also a pre-requisite for algorithms controlling the movements and
relative positions of drones in a swarm, e.g.\ algorithms based on
virtual spring models
\cite{Derr:Manic:11,Trotta:Montecchiari:DiFelice:Bonioni:20}. When it
comes to determining the movements of the swarm as a whole, this can
either be done in a distributed fashion (e.g., in the virtual spring
algorithm proposed in \cite{Derr:Manic:11}, the drones closer to a
location of interest ``pull'' the other drones towards that
location), or in a centralized ``leader-follower'' fashion, with one
``leader'' drone disseminating instructions to the other swarm members
\cite{Pantelimon:Tepe:Carriveau:Ahmed:17,Chen:Liu:Guo:20,Aggravi:Pacchierotti:Giordano:21}.

These various safety and coordination tasks require wireless
communication between drones. In this context, a drone swarm can be
regarded as a (potentially) large-scale, multi-hop and connected
wireless network. Currently, one major research strand is to use
modern cellular networks such as 4G, 5G or the upcoming 6G standard to
support communications between drones
\cite{Mishra:Natalizio:21,Geraci:GarciaRodriguez:Azari:Lozano:EtAl:22}. However,
these are not always available, in particular in disaster scenarios or
sparsely populated areas. Therefore, we are interested in employing
ad-hoc networking techniques for UAV communications, which often comes
under the heading of flying ad-hoc networks or FANETs
\cite{Baltaci:Dinc:Ozger:Alabbasi:Cavdar:Schupke:21,Chandrasekharan:Gomez:AlHourani:Kandeepan:Rasheed:Goratti:Reynaud:Grace:Bucaille:Wirth:Allsopp:16,Oubbati:Atiquzzaman:Lorenz:Tareque:Hossain:19,Chikri:Touati:Snoussi:Kamoun:19}.

The key goal of this paper is to provide networking support for
coordination and safety tasks in large-scale flying ad-hoc networks
such as drone swarms and formations, with only modest requirements on
the underlying wireless technology. The focus on coordination- and
safety-related tasks has some impact on the design of such protocols,
as these tasks often only require small packet sizes and many periodic
transmissions. For example, to avoid collisions with neighboured
drones, it is natural to adopt a similar approach as in ad-hoc
vehicular communications, whereby each vehicle frequently broadcasts
its current position, speed, heading, and other operational data to
neighboured vehicles, using a local broadcast. The overall size of
this operational data is quite limited, e.g.\ the required part of a
WAVE basic safety message transmitted in DSRC is in the order of 40
bytes \cite[Table 8]{Kenney:11}. Many road safety applications require
that vehicles transmit such messages at a rate of 10\,Hz
\cite{Karagiannis:Altintas:Ekici:Heijenk:Jarupan:Lin:Weil:10}. In this
paper we refer to the recurring transmission of such safety-related
data to a local neighbourhood as \define{beaconing}, and one such
locally broadcast packet is simply called a \define{beacon}. Another
motivating example for this work is centralized position and pose
control of a swarm or rigid drone formation
\cite{Aggravi:Pacchierotti:Giordano:21}. A rigid formation can be
described with only six degrees of freedom, three position and three
angular coordinates. If a drone formation operates in such a way that
each drone can derive its own trajectory from the knowledge of the
movements of the formation and its own position relative within the
formation, then it is sufficient for a leader drone to disseminate
timestamped records containing six floating point numbers into the
formation. These records should be disseminated quickly and reliably,
to achieve consistent trajectories and avoid collisions.

Different protocol mechanisms have been used in the literature to
support this kind of global dissemination of data, e.g.\ based on a
spanning tree construction, clustering or other topology-control
methods, or on flooding (see Section~\ref{sec:related-work}).  The
approach adopted in this paper rests \emph{entirely} on the periodic
local beaconing process.\footnote{A note on terminology: When we use
  the term periodic in this paper we do not mean strictly periodic,
  but allow for some variation in inter-beacon transmission times. We
  assume that there exists a well-defined average interval, and we
  refer to this average as the period.} The beacons are not only used
to transmit safety-related information to immediate neighbours, but we
also piggyback global coordination data and make sure that this data
is disseminated through the entire network. Specifically, we introduce
the abstraction of a globally shared \define{variable}, discussed
below. Piggybacking global coordination data onto beacons can reduce
packet overheads, as small pieces of coordination data are not wrapped
into their own packets anymore, but may come at the cost of increasing
packet length and delay. Furthermore, if the underlying medium access
control (MAC) protocol is based on CSMA (as is the case in WiFi
\cite{IEEE:802.11:16}), then piggybacking reduces the number of
channel access cycles --- in particular, the safety-critical beacon
packets do not have to compete anymore with other packets flooded into
the network for global dissemination. A further advantage of only
relying on locally broadcast beacons is that the dissemination does
not require routing or any other topological state (no parent-child or
similar routing relationships need to be maintained), and will
therefore be quite immune to any topology changes coming from
mobility, as long as connectivity is maintained. We also remark that
by relying on only one service of the underlying wireless technology,
the proposed approach is largely agnostic of this technology.

In this paper we describe and analyse a protocol stack built on these
ideas, the Drone Coordination Protocols (DCP) stack. At its base, DCP
enables periodic beaconing. To support global coordination and
collaboration in a swarm, one of the component protocols in our stack,
the VarDis (Variable Dissemination) protocol, offers the abstraction
of a globally shared \define{variable}, which can be dynamically
created, read, updated and deleted (CRUD). The protocol disseminates
information about any modifying operation on a variable (create,
update, delete) to the entire network within (repeated) beacon
payloads. It includes a mechanism by which nodes can recognize that
they do not have the most recent value of a variable, and it allows
such nodes to request that value from neighbours.  Every node that has
received a new modifying instruction for a variable (create, update,
delete) will repeat that instruction in a number of distinct
beacons. This approach combines repetition coding (or time diversity)
and, importantly, spatial diversity, as the same information is
repeated by several spatially distributed nodes. In more detail, in
this paper we make the following contributions:
\begin{itemize}
\item We describe the DCP (Drone Coordination Protocols) protocol
  stack, which includes at its base a simple protocol allowing to
  embed different payload types into frequently transmitted beacons
  (multiplexing), and the VarDis protocol, implementing the variable
  abstraction.
\item We present the results of a (mostly) simulation-based
  performance evaluation of VarDis, focusing on its reliability and
  latency in a range of scenarios. Our performance evaluation includes
  a sensitivity analysis, showing how performance depends on key
  VarDis and beaconing parameters. We also include a comparison of
  VarDis delay performance against numerical results obtained from a
  discrete-time Markov chain model in a simplified setting, to
  validate the simulation model. Furthermore, we also compare VarDis
  against a flooding protocol, which does not rely on periodic
  beaconing but floods each variable update separately.
\end{itemize}
Our results confirm that VarDis is a viable solution for global
dissemination of coordination data, particularly in more dense
networks. It is possible to tune its parameters to achieve quite
competitive delay and reliability performance, both in absolute
numbers and compared to a flooding protocol.

The protocol specification of the DCP stack and the VarDis protocol in
Version 1.0 is publicly available, as is the source code of the
simulation model used in this
paper.\footnote{\url{https://github.com/awillig/dcp-vardis.git}}

The remaining paper is structured as follows: we next discuss related
work in Section~\ref{sec:related-work}. In
Section~\ref{sec:dcp-vardis-description} we describe the DCP protocol
stack and the VarDis protocol in particular. Our system model and
setup for the simulation-based performance evaluation are discussed in
Section~\ref{sec:system-model}. In Section~\ref{sec:results} we
present our performance results for a range of scenarios. Conclusions
and a discussion of future work are given in Section
\ref{sec:conclusions}.

\section{Related Work}
\label{sec:related-work}

\subsection{Data Dissemination}
\label{subsec:related-work:data-dissemination}

Many different protocols have been used to support global data
dissemination in a wireless network.

In one broad class of protocols suitable routing structures are
established. For example, in \cite{Chen:Liu:Guo:20} a spanning tree
rooted in the leader is constructed based on hop distances to the
leader node.  Each follower picks a suitable parent and retrieves the
required information from it. A similar example from the sensor
network domain is the CTP protocol \cite{TinyOS:tep123:06},
\cite{Gnawali:Fonseca:Jamieson:Moss:Levis:09}. Other examples are
based on clustering \cite{Ucar:Ergen:Ozkasap:16,Li:Dai:09}.  In the
presence of significant relative mobility, however, such protocols
have to spend some effort and time on maintaining and updating these
routing structures, and there is always a risk of stale
entries.

Approaches based on opportunistic or epidemic data dissemination such
as the one proposed in
\cite{Wu:Arkhipov:Kim:Talcott:Regan:McCann:Venkatasubramanian:17} work
naturally in intermittently connected networks, but often only address
the case of unicast delivery and do not guarantee that a given piece
of information is provided to \emph{every} node in the network. They
also differ from our semantics of a variable --- usually they treat
different messages as entities to be disseminated individually and
completely, and do not allow to treat one message as an updated
version of another one and overwrite the old version.  Finally, the
context in which opportunistic or epidemic protocols are analyzed is
often very different from the context of online control with tight
latency targets relevant for swarm control. Another approach is to
rely on classical flooding or optimized flooding techniques
\cite{Alshaer:Horlait:05} --- in a lightly loaded network these indeed
can disseminate information to everyone quickly and with very high
reliability, but they have a lot of overhead (particularly when the
data items to be disseminated are small), they are susceptible to the
broadcast storm problem \cite{Tseng:Ni:Chen:Sheu:02}, and can become
very inefficient when several flooding operations triggered by
different nodes run in
parallel.

Perhaps closest in spirit to our designs are gossip protocols as used
in consensus schemes
\cite{Bakhshi:Cloth:Fokkink:Haverkort:08,Chaintreau:LeBoudec:Ristanovic:09,Aysal:Yildiz:Sarwate:Scaglione:09,Shi:Li:Johansson:Johansson:16,Dhuli:Gaurav:Singh:15,Jiang:Cao:Krishnamachari:Zhou:Niu:20,OlfatiSaber:Murray:04}.
In consensus problems, nodes attempt to reach agreement on a
particular value, for example the average temperature. Each node
starts with its own local temperature observation, collects
observations from neighbours and processes these together with its own
current value to calculate a new estimate, and then in turn transmit
this estimate to neighbours. If implemented correctly, the local
estimates for nodes will converge to the average.  A range of link
models and node failure models are varied in these investigations,
often the analysis focuses on (time until) convergence, e.g.\ when
calculating the average of the initial values or their maximum
\cite{Muniraju:Tepedelenlioglu:Spanias:19,Iutzeler:Ciblat:Jakubowicz:12,Molinari:Agrawal:Stanczak:Raisch:21}. In
terms of protocols, consensus schemes often either use global
broadcasts (transmissions of nodes with sufficient transmit power to
reach every other node) or gossiping, where information is
disseminated through the network using local broadcasts. It is
important to note that the semantics of consensus problems differs
from the semantics of a variable in VarDis.

\subsection{Congestion Control}
\label{subsec:related-work:congestion-control}

In our simulations, we use beaconing on top of IEEE~802.11g. The
overall setup is quite similar to the beaconing process known from
vehicular networks. The beacons are broadcast locally and
unacknowledged, and since the most-used IEEE~802.11g MAC protocols
(DCF or EDCA) are CSMA-based, they suffer from direct or
hidden-terminal collisions, leading to packet losses. This problem,
which is often referred to as congestion control problem, is
well-studied
\cite{Shen:Cheng:Zhang:Jiao:Yang:13,Sahoo:Wu:Sahu:Gerla:13,Hassan:Vu:Sakurai:11,Yao:Rao:Liu:13},
and some of the results presented in this paper are affected by this
problem.

\section{The Drone Control Protocol (DCP) and the Variable
  Dissemination (VarDis) Protocol}
\label{sec:dcp-vardis-description}

In this section we describe the overall DCP protocol stack and the
variable dissemination (VarDis) protocol.

\subsection{Overall Architecture}
\label{subsec:dcp-vardis-description:overall-architecture}

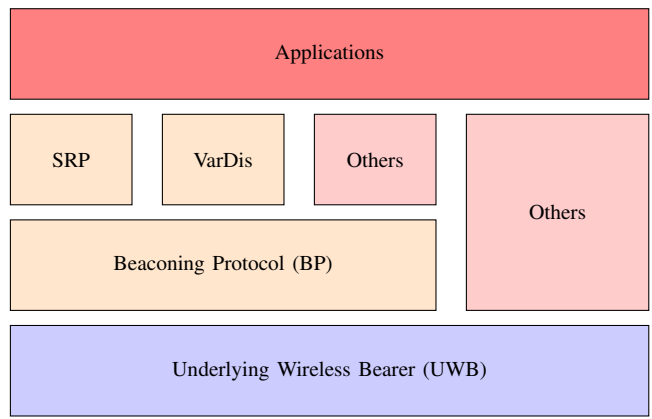
\begin{figure}
  \centering
  \begin{tikzpicture}[scale=0.4]
    \filldraw[fill=red!50!white, draw=black] (0,10.5) rectangle (21,13.5) node[pos=.5]{\footnotesize{Applications}};
    \filldraw[fill=red!20!white, draw=black] (15,3.5) rectangle (21,10) node[pos=.5]{\footnotesize{Others}};
    \filldraw[fill=orange!20!white, draw=black] (0,7) rectangle (4,10) node[pos=.5]{\footnotesize{SRP}};
    \filldraw[fill=orange!20!white, draw=black] (5,7) rectangle (9,10) node[pos=.5]{\footnotesize{VarDis}};
    \filldraw[fill=red!20!white, draw=black] (10,7) rectangle (14,10) node[pos=.5]{\footnotesize{Others}};
    \filldraw[fill=orange!20!white, draw=black] (0,3.5) rectangle (14,6.5) node[pos=.5]{\footnotesize{Beaconing Protocol (BP)}};
    \filldraw[fill=blue!20!white, draw=black] (0,0) rectangle (21,3) node[pos=.5]{\footnotesize{Underlying Wireless Bearer (UWB)}};
  \end{tikzpicture}
  \caption{DCP Protocol Stack}
  \label{fig:dcp-protocol-stack}
\end{figure}

The overall DCP protocol stack and its environment are shown in
Figure~\ref{fig:dcp-protocol-stack}.  At the lowest layer, and outside
the DCP protocol stack, we have the \define{underlying wireless
  bearer} (UWB). We do not make strong assumptions about its abilities
and services, we only assume that it provides a local broadcast
service, i.e.\ it allows a node to broadcast a packet to a local
single-hop neighbourhood (generally determined by transmit power,
antenna characteristics and propagation environment), and to receive
and process such broadcasts from neighboured nodes. We also assume
that it allows for a maximum packet size of at least 200 -- 300
bytes. The local broadcasts are generally unacknowledged, but the
packets are protected by checksums, and packets with bit errors are
dropped.  Based on this and on the general absence of receiver
feedback for local broadcasts, we have chosen not to include separate
error detection or correction mechanisms into any of the DCP
protocols.  We also assume that the UWB provides protocol
multiplexing, so that it can run DCP in parallel to other client
protocols such as a TCP/UDP/IP stack. For our performance evaluation
the UWB is provided by WiFi \cite{IEEE:802.11:16}, such that DCP
packets are directly encapsulated into WiFi payloads.

The lowest layer of DCP proper is provided by the \define{beaconing
  protocol} (BP), which operates on top of the UWB. The beaconing
protocol is mainly responsible for the frequent generation and
transmission of \define{beacons} including one or more \define{client
  payloads}, coming from BP client protocols such as the State
Reporting Protocol (SRP) or VarDis. The beacons are transmitted using
the local broadcast service of the UWB. The beacon generation rate is
configurable.  In general, the protocol does not specify which
payloads from client protocols are included in any specific beacon,
but for the purposes of this paper we only include VarDis as a client
protocol, and VarDis payloads consequently are always included. In the
receive direction, the BP can receive beacons sent by neighboured
nodes and hand over the included payloads to their respective client
protocols. Client protocols can dynamically register and de-register
with BP, and are being offered different interfacing methods (e.g.\
queues vs.\ single-packet buffers).  In other words, BP provides
protocol multiplexing amongst client protocols. The BP is described in
more detail in Section
\ref{subsec:dcp-vardis-description:beaconing-protocol}.

The State-Reporting Protocol (SRP) is a client protocol of BP. Its
main responsibility is to periodically generate payloads containing
the position, speed, heading and other operational data of the sending
drone. This data is only sent to immediate neighbours and its main
purpose is the avoidance of drone collisions. In the receive
direction, an SRP entity uses received SRP payloads to build a
\define{neighhbour table}, which lists all the currently known
neighbours, their positions / speed / heading etc., and the time that
has passed since the last reception of a SRP payload. The main purpose
of the neighbour table is to allow a drone to assess the uncertainty
in a neighbours position and predict impending collisions. The entries
of the neighbour table use a soft-state mechanism to push out stale
entries. We will not discuss the SRP any further, as it is outside the
scope of this paper.

The Variable Dissemination (VarDis) protocol also sits as a client
protocol on top of BP. It allows applications to create, read, update
and delete (CRUD) globally shared \define{variables}, which VarDis
disseminates into the entire network, with the goal to achieve a
consistent view after any modifying variable operation (create,
update, delete) as quickly and reliably as possible. In the VarDis
version reported in this paper (Version 1.0), update and delete
operations are restricted to the node that created the variable.  The
protocol also contains mechanisms for nodes to detect missing
variables or updates, and to request this missing information from
neighbours. VarDis is described in more detail in Section
\ref{subsec:dcp-vardis-description:vardis}.

\subsection{Beaconing Protocol (BP)}
\label{subsec:dcp-vardis-description:beaconing-protocol}

As mentioned above, the key tasks of BP are:
\begin{inparaenum}[(i)]
\item the periodic transmission of beacons carrying client protocol
  payloads to the local neighbourhood, using local broadcasts;
\item the reception of beacons from neighbours and forwarding received
  client protocol payloads to their respective client protocol; and
\item the dynamic registration, query and de-registration of these
  client protocols.
\end{inparaenum}

In the transmit direction, the interface between a client protocol and
BP offers three different ways of buffering payloads:
\begin{itemize}
\item Queueing: payloads handed over by the client protocols are
  stored in a first-in-first-out queue of indefinite size. Once a
  payload has reached the head of the queue, it can be inserted into
  an outgoing beacon. After its insertion the payload is
  dropped. Hence, each payload is transmitted exactly once.
\item Buffered-Once: payloads are stored in a single buffer and can be
  overwritten by new payloads if they have not been transferred into a
  beacon yet. Once the buffer contents have been transferred into a
  beacon, the buffer is cleared. Hence, each payload is transmitted at
  most once.
\item Buffered-Repeated: payloads are stored in a single buffer and
  can be overwritten by new payloads. In contrast to Buffered-Once,
  the buffer is never cleared, i.e.\ the same payload is inserted into
  subsequent beacons as long as it is not overwritten. Hence, each
  payload can be transmitted an arbitrary number of times, including
  zero.
\end{itemize}
The protocol specification leaves it to implementations to define at
which times the BP generates new beacons, it only stipulates that the
long-term average beacon transmission rate converges to a configurable
value (this configuration can be changed at runtime). For the purposes
of this paper, we assume that the inter-beacon generation times are
independent and identically distributed (iid) random variables, which
has the benefit of de-synchronizing the nodes. In our simulations,
the beacon transmissions occur with an average rate of $\beta$ beacons
per second, where the inter-beacon times are either taken from an
exponential distribution with rate parameter $\beta$, or from
$U\left[\frac{0.9}{\beta},\frac{1.1}{\beta}\right]$, which we refer to
as ``periodic with 10\,\% jitter'' at rate $\beta$ in the remainder of
this paper.

The protocol specification also does not prescribe in which order the
buffers or queues associated with the different client protocols are
checked to see whether they have a payload ready, this is left to
implementations (these should generally make sure that each beacon
contains an SRP payload when the SRP protocol is present). However, in
this paper the only client protocol considered is VarDis.

\subsection{Variable Dissemination Protocol (VarDis)}
\label{subsec:dcp-vardis-description:vardis}

VarDis\footnote{The specification is available at
  \url{https://github.com/awillig/dcp-vardis.git}.} is a BP client
protocol utilizing the Buffered-Once queueing mode. On the highest
level, VarDis maintains a \define{real-time database}\footnote{Our use
  of the term ``real-time'' does not imply that we can guarantee any
  latency or reliability, it only represents our aspiration to achieve
  low latency and high reliability when disseminating changes to
  variable values.}, which is a collection of globally shared
\define{variables}.  A variable refers to something which has a unique
identifier (represented by an integer value), a current value
generally varying over time (represented by a length value and a
contiguous block of bytes without any further discernible structure),
and an indication of how recent it is (represented through sequence
numbers). The protocol aims to disseminate all changes to the variable
value to all drones in as little time and as reliably as possible. The
byte blocks are assumed to be of small to moderate size (significantly
smaller than the maximum beacon size), VarDis does not support a
fragmentation-and-reassembly type of mechanism for individual
variables. We support four basic operations on a variable: creating
it, reading its value, updating its value, and deleting it
(CRUD). Variables can be created and deleted dynamically, their
variable identifier must be unique.\footnote{Uniqueness must be
  ensured by the users, it is not strongly enforced by the
  protocol. In particular, when two nodes multiple hops away create a
  variable with the same variable identifier at about the same time,
  some nodes in the middle will be able to detect this condition, but
  the protocol does not contain a mechanism to repair it or to even
  signal this condition.}

Currently, each variable in VarDis has one \define{producer} node: the
producer node is the unique node creating the variable, and this node
is also the only node allowed to update the value of the variable or
deleting it --- in other words, our variables are single-writer. All
other nodes are only allowed to read the value of the variable, we
occasionally also refer to them as
\define{consumers}.\footnote{Conceptually, VarDis is not a
  publish/subscribe scheme
  \cite{Eugster:Felber:Guerraoui:Kermarrec:03}, as every node consumes
  every variable, without there being a choice concerning its
  subscription.}

\subsubsection{Variable Attributes}
\label{subsubsec:dcp-vardis-description:variable-attributes}

A variable is described by its producer through a \define{variable
  specification}, which includes its variable identifier
(\texttt{varId}), the unique node identifier (e.g.\ MAC address) of
the variable producer, a textual description of the variable, and a
repetition counter (\texttt{repCnt}), which specifies how often each
node should repeat an instruction to create, update, or delete a
variable in distinct beacons.

The producer of a variable furthermore maintains a sequence number for
it, which is initialized to zero at the time of its creation, and
incremented for each new update operation. When a variable update is
disseminated through the network, the sequence number is included so
that consumer nodes can check whether a received update is truly more
recent than what they currently have stored in their local real-time
database.

Any consumer node\footnote{Which includes the producer as well, which
  additionally is a consumer of its own variable.} maintains a
\define{database entry} for a variable, with its \texttt{varId} as a
key. This database entry contains the variable specification, the
current variable value (both length and byte array), the sequence
number of the last received update, a local timestamp representing the
time this last update was received, and counters telling how many
further repetitions of a variable create / update / delete instruction
are to be included in future distinct beacons. The operation of
reading a variable simply refers to retrieving its current value from
its database entry. A consumer can furthermore retrieve a list of
current database entries from the database.

\subsubsection{Variable Creation, Update and Deletion}
\label{subsubsec:dcp-vardis-description:modifying-operations}

When the application on the producer node of a variable decides to
create that variable, it hands over a variable specification and an
initial value for the variable. The producer wraps both into a data
structure signifying an instruction to create a variable to other
nodes. It then includes this data structure into a given number of
distinct beacons (c.f.\ the \texttt{repCnt} parameter). Each other
network node receiving this instruction for the first time adds the
new variable to its real-time database, and repeats the create
instruction in as many future distinct beacons as given by the
\texttt{repCnt} parameter.

The same approach of having all other network nodes repeat an
instruction in as many future distinct beacons as indicated in the
\texttt{repCnt} parameter of a variable is also followed when the
producer disseminates the instruction to delete a variable, or when it
disseminates an update to a variable. Such an update instruction
carries the \texttt{varId} of the variable, its current sequence
number (\texttt{seqno}) as assigned by the producer, and its new
length and value. When a consumer receives an update instruction for
an existing variable, it will:
\begin{itemize}
\item Discard the update when it has the same sequence number as
  stored in the local real-time database.
\item Update the local real-time database when the received update
  instruction has a strictly more recent sequence number than stored
  in the real-time database.
\item Discard the update when it has a strictly older sequence number
  than stored in the local real-time database, and additionally
  schedule \texttt{repCnt} transmissions of the current variable value
  (in an update instruction) in distinct future beacons. The purpose
  of this is to correct an outdated value in the original sender of
  the update instruction.
\end{itemize}
In this paper we ignore issues around the representation of sequence
numbers, wrap-overs and re-start after node crashes.

\subsubsection{Summary Mechanism, Update and Create Requests}
\label{subsubsec:dcp-vardis-description:summary-mechanism}

If there is sufficient space in a beacon, each node includes
\define{summaries} of the variables in its internal real-time database
into its beacons. Generally not all variables can be summarized in a
single beacon, so the summaries are sent in a round-robin fashion. A
summary for a variable includes its \texttt{varId} and its
\texttt{seqno}. The summaries sent by a node are not repeated by its
neighbours.

When a node $A$ receives a variable summary from a neighbour, it first
checks the \texttt{varId}. When $A$ does not have that variable in its
local real-time database, it includes a special request instruction in
one of its next beacons. This request instruction (``create-request'')
carries the \texttt{varId} of the missing variable and any neighbour
receiving this will include a variable-creation instruction in as many
future distinct beacons as the \texttt{repCnt} for this variable
indicates. This mechanism gives $A$ a way to learn the specification
and current value of a variable it did not previously have.

If on the other hand $A$ does have the variable in its real-time
database, but in a strictly outdated version (as recognizable from the
sequence numbers), then $A$ includes another special request
instruction in a future beacon (``update-request''). In this
update-request instruction, $A$ includes the \texttt{varId} and the
sequence number from its real-time database. Any neighbour hearing
this update-request will include variable updates into \texttt{repCnt}
future distinct beacons, provided they have strictly more recent
updates than the requesting node.

It is important to note that VarDis does not guarantee that a node
will receive every update ever issued by the producer --- if updates
are generated too quickly, newer values may ``over-write'' older ones
during dissemination. This leads to ``update gaps'', which we
investigate in our performance evaluation.

\subsubsection{Packet Construction and Transmit Path}
\label{subsubsec:dcp-vardis-description:transmit-path}

A VarDis beacon payload can have up to six different sections. These
sections are prioritized amongst each other, and a section is only
added to a beacon when there is sufficient remaining space (VarDis
allows to limit its maximum payload size). The highest-priority
section are the instructions to create variables, followed by
instructions to delete variables. Third are variable updates, then
variable summaries. Fifth priority have create-request instructions,
and finally then update-request instructions are being
included. Within each section (except the section for variable
summaries), entries for variables are added as long as space is
available. The number of summaries to be included in the summary
section is upper-bounded by a configurable protocol parameter,
referred to as \texttt{maxSumCnt} in the remainder of the paper.

VarDis provides payloads to the underlying BP for every outgoing
beacon. If all sections are empty, then no payload is being prepared.

\section{System Model and Simulation Setup}
\label{sec:system-model}

We primarily use a simulation-based approach to explore the
performance of VarDis and BP in different scenarios and for different
parameter combinations.  For our simulations we use WiFi
(IEEE~802.11g) with the distributed coordination function (DCF) as the
UWB. All simulations were carried out using OMNeT++ in version
6.0\footnote{\url{https://www.omnetpp.org}} and the INET framework in
version 4.4\footnote{\url{https://inet.omnetpp.org}}, which provides
the models for the IEEE 802.11g radio and channel. Our simulation code
and setup files are available
online\footnote{\url{https://github.com/awillig/dcp-vardis.git}. It
  should be noted that the simulation code does not fully conform to
  the specification, but the differences are irrelevant for the
  purposes of this paper.}

In this section we describe our system model in more detail. In
Table~\ref{tab:fixed-parameters} we summarize some key parameters that
we keep fixed throughout, whereas Table~\ref{tab:varied-parameters}
describes the parameters we vary. We also define a simple flooding
protocol to compare VarDis against.

\begin{table}
  \begin{center}
    \begin{scriptsize}
      \begin{tabular}{p{1cm}p{5cm}l}
        \toprule
        \textbf{Parameter} & \textbf{Explanation} & \textbf{Value} \\
        \midrule
        \multicolumn{3}{l}{\textbf{Channel and PHY parameters}} \\
        $\eta$        & Path loss exponent & 2.25 \\
                           & PHY data rate  & 12\,Mbps \\
        $PL_0$          & Path loss at reference distance 1\,m & 40\,dB \\
        $N_0$           & Thermal noise intensity & $-174$\,dB/Hz \\
                           & PHY Data rate (code rate 3/4, 16-QAM) & 36\,Mb/s \\
                           & Transmit power & 0\,dBm \\
        \midrule
        \multicolumn{3}{l}{\textbf{BP and VarDis parameters}} \\
                        & Maximum beacon length & 200\,B \\
                        & Random inter-beacon jitter & 10\,\% \\
                        & Variable length & 12\,B \\
        \bottomrule
      \end{tabular}
    \end{scriptsize}
    \caption{Fixed Parameters}
    \label{tab:fixed-parameters}
  \end{center}
\end{table}

\begin{table}
  \begin{center}
    \begin{scriptsize}
      \begin{tabular}{p{1.3cm}p{5cm}}
        \toprule
        \textbf{Parameter} & \textbf{Explanation}  \\
        \midrule
        \multicolumn{2}{l}{\textbf{Deployment Parameters}} \\
        $P$        & Packet error rate on links (\%) \\
        $K$        & Controls number of nodes in deployment ($K$ in
                     line deployments, $K^2$ in grid deployments) \\
        \midrule
        \multicolumn{2}{l}{\textbf{BP and VarDis parameters}} \\
        $\beta$            & Beacon transmission rate (Hz) \\
        $\lambda$          & Average inter-arrival time for variable
                             updates at a producer (s) \\
        \texttt{repCnt}    & repetition counter for modifying
                             operations\\
        \texttt{maxSumCnt} & Maximum number of summaries in a beacon\\
        \bottomrule
      \end{tabular}
    \end{scriptsize}
    \caption{Varied parameters}
    \label{tab:varied-parameters}
  \end{center}
\end{table}

\subsection{Channel Model and Physical Layer Settings}
\label{subsec:system-model:channel-phy}

To allow us to focus purely on the effect of VarDis protocol
mechanisms, we opt for a relatively simple air-to-air channel model
that takes path loss into account, but does not model shadowing,
fading or external interference. While idealized, these settings
correspond to a situation where a drone swarm flies high enough and
uses a sufficiently small transmit power to inhibit multipath
propagation. In particular, we employ the log-distance path loss model
\cite{Rappaport:02} with path loss exponent $\eta=2.25$ and path loss
at reference distance $PL_0=40$ dB (the reference distance is
1\,m). The only noise source is thermal noise at $-174$\,dBm/Hz.

On the physical layer we use IEEE 802.11g settings, operating in the
2.4\,GHz band (channel 1, centre frequency of 2,412\,MHz), using OFDM
with a 36\,Mb/s modulation and coding scheme (16-QAM, 3/4 code rate),
and a transmit power of 0\,dBm. The simulated radio has a receiver
sensitivity of $-96$\,dBm. With these settings the transmission range
(interpreted as the distance within which packet reception is still
possible with a packet error rate of less than 10\,\%)\footnote{All
  references to packet error rates made in this paper are with
  reference to packets with a 50B MAC payload, corresponding to 77B
  bytes transmitted in total.} is 255\,m, whereas the interference
range (defined as the distance at which the receiver judges the signal
energy to just still be above the carrier-sense threshold, but without
being able to decode successfully) is 294\,m.

\subsection{Deployments and Traffic}
\label{subsec:system-model:deployment-traffic}

\begin{figure}
  \centering
  \resizebox{0.75\columnwidth}{!}{\includegraphics{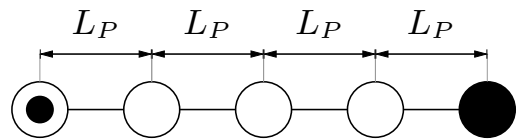}}
  \caption{Line deployment with fixed density and $K$ nodes. The
    half-filled node is the sole producer of a variable, the fully
    filled node is the reference consumer. Each link is $L_P$ meters,
    where $P$ is a given packet error rate.}
  \label{fig:line-deployment:fixed-density}
\end{figure}

\begin{figure}
    \centering
    \resizebox{0.75\linewidth}{!}{\includegraphics{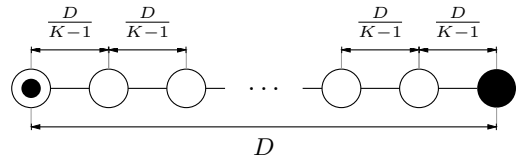}}
    \caption{Line deployment with variable density and $K$ nodes. The
    half-filled node is the sole producer of a variable, the fully
    filled node is the reference consumer.}
    \label{fig:line-deployment:variable-density}
\end{figure}

We investigate node deployment scenarios of different density and
size, but in all scenarios the nodes are stationary. We simulate two
different types of line deployments with $K$ nodes each.  In
\define{fixed-density} line deployments, we place nodes at a fixed
distance $L_P$ between neighbours, where $L_P$ is chosen to achieve a
given packet error rate of $P$\,\% over one link --- compare
Figure~\ref{fig:line-deployment:fixed-density}. In
\define{variable-density} line deployments (cf.\
Figure~\ref{fig:line-deployment:variable-density}) we also place nodes
equidistantly, but we keep the position of the two end nodes fixed to
$D=1,120\,\mbox{m}$ as $K$ (and with it the inter-node distance)
varies.

\begin{figure}
    \centering
    \resizebox{0.75\linewidth}{!}{\includegraphics{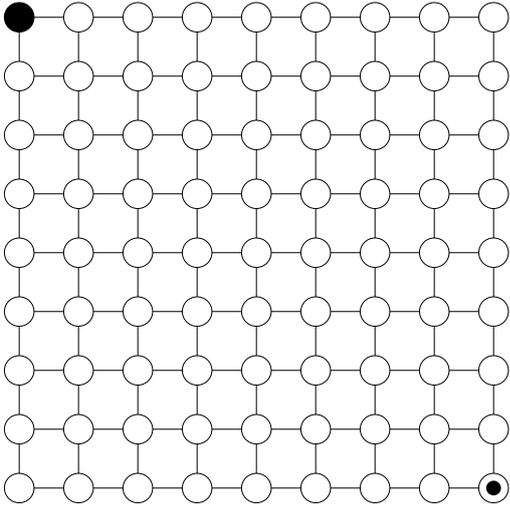}}
    \caption{Grid deployment with $K$ nodes on a side and $K^2$ nodes
      in total, with either fixed density (all vertical or horizontal
      links have common packet error rate $L_P$) or variable density
      (total side length fixed to 1,120\,m). All nodes produce a
      variable, the producer and consumer we focus on are highlighted
      as the half-filled node in the lower right corner (producer) and
      the fully filled node in the upper left corner (consumer).}
    \label{fig:grid-deployment}
\end{figure}

We also simulate two different types of regular grid deployments, with
$K$ nodes on each side and $K^2$ nodes in the overall deployment (cf.\
Figure~\ref{fig:grid-deployment}). In fixed-density grid deployments
the length of one vertical or horizontal link is fixed to $L_P$, to
achieve a given packet error rate $P$\,\% on that link, whereas in a
variable-density deployment the total side length (from one corner
node to a neighboured corner node) is kept fixed at 1,120\,m as $K$
varies.

For our simulations, we created a dummy VarDis application and run one
instance on every node. The nodes chosen as producers update their
variable with an average inter-update time of $\lambda$\,s, which is
either strictly periodic (in the line experiments) or follows an
exponential distribution (in the grid experiments). Each of these
variables is 12\,B long, with 8\,B used to encode the simulation time
at which the variable was updated at the producer (as a double float),
and 4\,B are used for an application sequence number.\footnote{Using a
  separate sequence number in an application variable allows to make
  it longer than the sequence numbers we use in a VarDis variable
  update, and removes the need to instrument the VarDis code as such.}
Each producer initializes its local sequence number to zero and
increments it before each variable update. The timestamps can be used
by a consumer node to measure the end-to-end application-layer update
delay (time between generation of the update at the producer and its
reception at the consumer), and the sequence number can be used to
identify whether the consumer has missed any updates. The update delay
and the difference in received sequence number (``gap'') are the two
metrics we use to judge the performance of VarDis in the rest of the
paper (cf.\ Section~\ref{subsec:system-model:performance-metrics}).

In the line deployments we have only one producer of a variable, and
we consider only one consumer --- these two nodes are placed on either
end of the line. In the grid deployments, \emph{all} nodes produce a
variable, but when showing our results we focus on one particular
producer (lower right node in the grid, see
Figure~\ref{fig:grid-deployment}) and one particular consumer (upper
left node). We consider these two reference nodes to represent the
``worst-case''.

\subsection{MAC Protocol and Beaconing}
\label{subsec:system-model:mac-beaconing}

On the MAC layer we use the IEEE 802.11 DCF with default contention
window parameters. The network runs in ad-hoc mode without any
requirements for authentication or association. We only generate
beacons containing VarDis payloads, there is no other traffic source
present in the system. The beacons are locally broadcast.  Throughout
all simulations the maximum beacon size is set to 200\,B (except where
noted otherwise), SRP payload generation is disabled. The beacon
inter-arrival times are drawn independently from the uniform
distribution over $\left[\frac{0.9}{\beta},\frac{1.1}{\beta}\right]$
(we sometimes refer to this as ``periodic with 10\,\% jitter'') or
from an exponential distribution with rate parameter $\beta$. In both
cases, $\beta$ (in Hz) is the beacon generation rate. It is a variable
parameter.

\subsection{Variable Parameters}
\label{subsec:system-model:variable-parameters}

There is a number of parameters which we vary directly or indirectly
in our performance evaluation: the number of drones (through varying
$K$), the drone density (through varying $K$ in variable-density
settings), the network diameter (through $K$ in fixed-density
settings), the beaconing rate $\beta$, or the inter-arrival times
$\lambda$ for variable updates.  Furthermore, we consider different
settings for some key VarDis parameters: whether or not the
summarization mechanism is enabled, the value of the repetition count
parameter \texttt{repCnt}, and the maximum number of variable
summaries included in a beacon (\texttt{maxSumCnt}). The variable
parameters are summarized in Table~\ref{tab:varied-parameters}.

\subsection{Main Performance Metrics}
\label{subsec:system-model:performance-metrics}

In all simulations reported below we will have one selected producer
and one selected consumer of a variable as described in
Section~\ref{subsec:system-model:deployment-traffic}. We measure the
reception reliability for the consumer node by evaluating the average
application layer sequence number gap between successively received
variable updates (recall that the variable itself includes a sequence
number and timestamp, see Section
\ref{subsec:system-model:deployment-traffic}). In the absence of
update losses, that average gap would be exactly one, whereas a higher
value indicates losses of updates. Similarly, in the consumer node we
measure the end-to-end update delay by subtracting the timestamp in
the variable from the timestamp of reception at the consumer.

Error bars --- if shown --- represent the 95\%{} confidence intervals
of the result. Unless otherwise mentioned, for each parameter
combination we collected at least 100,000 variable updates at the
chosen consumer across a number of replications which use different
random number seeds. The in-simulation-time and replication count were
dependent on the variable update rate and the computing resources
available to us.

\subsection{Flooding Protocol}
\label{subsec:system-model:flooding-protocol}

As part of our experiments we compare the performance of VarDis
against a flooding protocol, in which each variable update triggers a
separate flooding operation. More specifically, an update for a
variable as described in
Section~\ref{subsec:system-model:deployment-traffic} is handed down to
the flooding protocol, which encapsulates it in its own packet. The
flooding protocol header is 14\,B long and includes its own sequence
number for duplicate detection, time-to-live (irrelevant for the
purposes of this paper), and source address fields. The flooding
protocol then puts the packet into a broadcast queue for transmission.

The flooding protocol has a separate worker process which constantly
pulls packets from this broadcast queue, waits a random backoff
period, hands over the packet to the MAC for transmission, and then
starts over. Such a backoff helps to reduce the number of collisions
between forwarder nodes that have received this update at the same
time. The backoff period is drawn from an exponential distribution
with an average back-off period of 10\,ms. The flooding protocol also
repeats transmissions, putting the last transmitted packet back at the
front of the queue until \texttt{repCnt} transmissions have occurred.

When a station receives a broadcast, it checks if it is the source of
the packet, or whether it has seen the same packet or one with a more
recent sequence number before. If either of these conditions are met,
it drops the packet. Otherwise, the station passes the message up to
the application and adds the packet to the broadcast queue for further
dissemination.

It should be noted that this flooding protocol is somewhat adapted to
(and optimized for) the concept of variables that we use in this
paper. In particular, a forwarder in a fully generic flooding protocol
would only drop received packets if they are exactly the same as
packets that have been forwarded recently (as determined from the node
address of the sender and a sequence number), and not drop it already
when it has seen a more recent sequence number. In other words, a
fully generic flooding protocol would not be able to tell ``younger''
from ``older'', it would only test for equality. Note that this
adaptation is beneficial for the flooding protocol, as it reduces the
number of packets transmitted. Furthermore, fully
generic flooding protocols would not repeat the same packet
\texttt{repCnt} times, as happens in VarDis.

\section{Results}
\label{sec:results}

In this section we discuss our results. We first consider
fixed-density networks. For these, we demonstrate the effect of
summarization for a line network, then study a grid network to perform
a sensitivity analysis for key DCP and VarDis parameters, and finally
compare VarDis against a flooding protocol. We then turn our attention
to variable-density networks. First, we consider again the case of the
line network. To at least partially validate our simulation model, we
compare simulation results against numerical results obtained from a
discrete-time Markov chain model. We then evaluate grid networks with
a substantially higher traffic load wherein each node produces and
frequently updates a variable, and present results showing how often
nodes can update their variables subject to certain delay and
reliability (average sequence number gap) constraints.

\subsection{Fixed-Density Networks: The Effect of Summarization}
\label{subsec:results:fixed-density:summarization-effects}

We consider the impact of summarization in a fixed-density line
network (Section~\ref{subsec:system-model:deployment-traffic}, in
particular Figure~\ref{fig:line-deployment:fixed-density}). We
consider per-hop packet error rates of
$P\in\set{20\,\%, 50\,\%, 80\,\%}$ and choose the link distance $L_P$
accordingly
($L_P\in\set{263\,\mbox{m}, 273\,\mbox{m}, 280\,\mbox{m}}$). In all
cases, the chosen settings imply that each drone will only be able to
receive packets from immediate neighbours, the distance to two- or
more-hop neighbours is too large to receive any packets. The sole
producer of a variable generates variable updates every 5\,s. We only
recorded delay and gap size results for the last drone in the chain,
below referred to as the consumer.

We have varied the following parameters: the status of VarDis
summarization (enabled or disabled), the VarDis repetition count
($\mathtt{repCnt}\in\set{1,2,3}$), the number of nodes in the line
($K\in\set{3,4,\ldots,17}$), and the beaconing rate
($\beta\in\set{10\,\mbox{Hz}, 20\,\mbox{Hz}}$).

\begin{figure*}
  \subfloat[With summaries]{\includegraphics[width=0.5\textwidth]{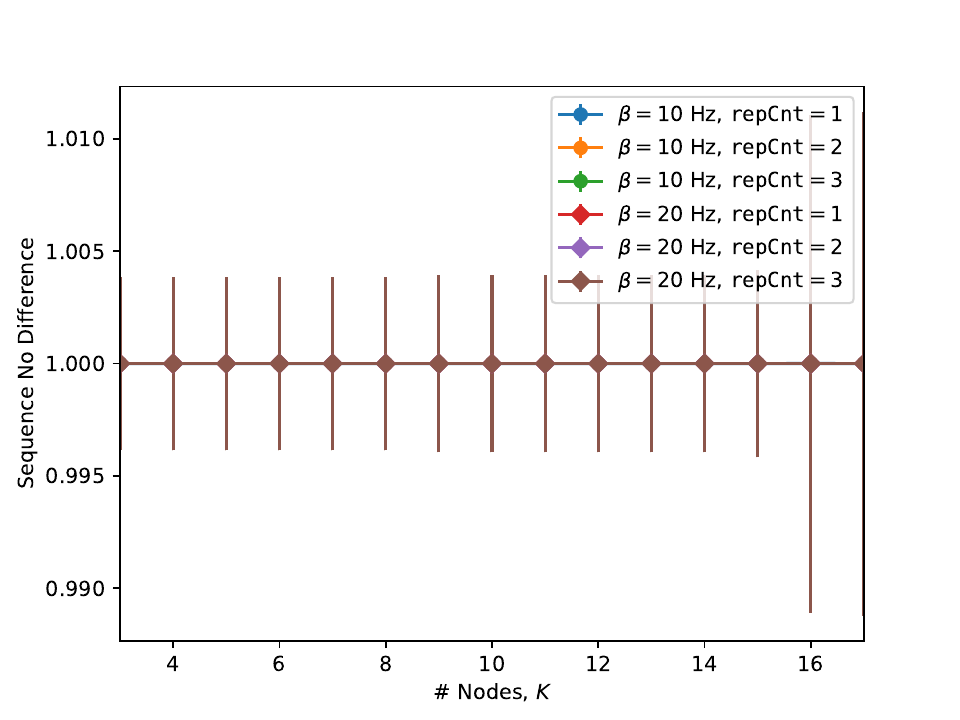}}
  \subfloat[Without summaries]{\includegraphics[width=0.5\textwidth]{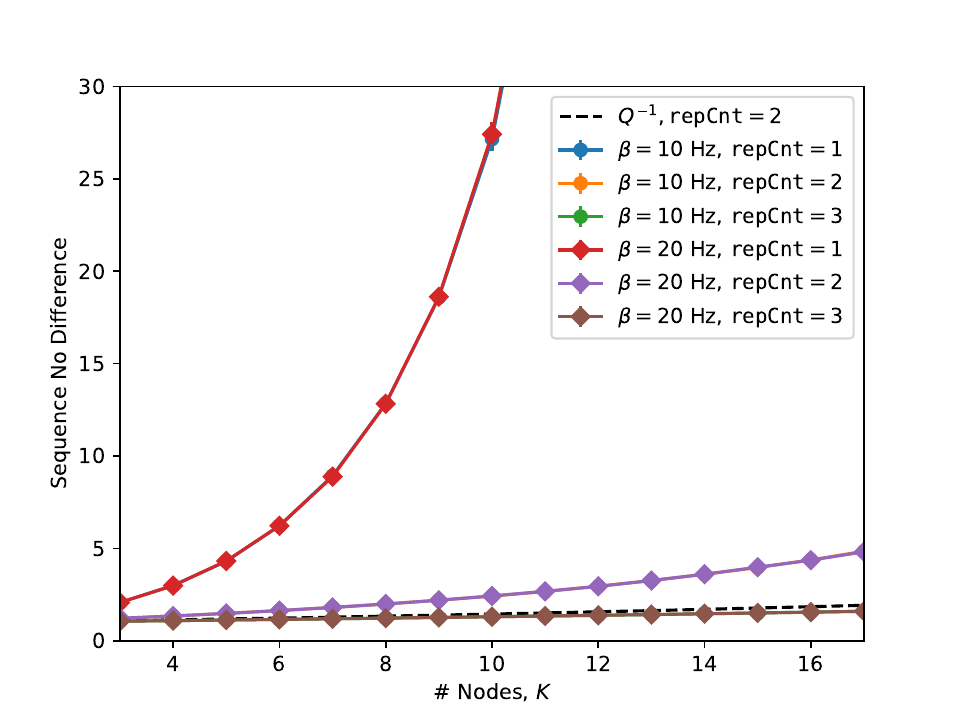}}
  \caption{Average sequence number gap for per-hop PER $P=20\,\%$,
    with and without summaries, varying \texttt{repCnt} and beaconing
    rate.}
  \label{fig:experiment-one-avg-seqno-gap-20-percent}
\end{figure*}

\begin{figure*}
  \subfloat[$P=50\,\%$]{\includegraphics[width=0.5\textwidth]{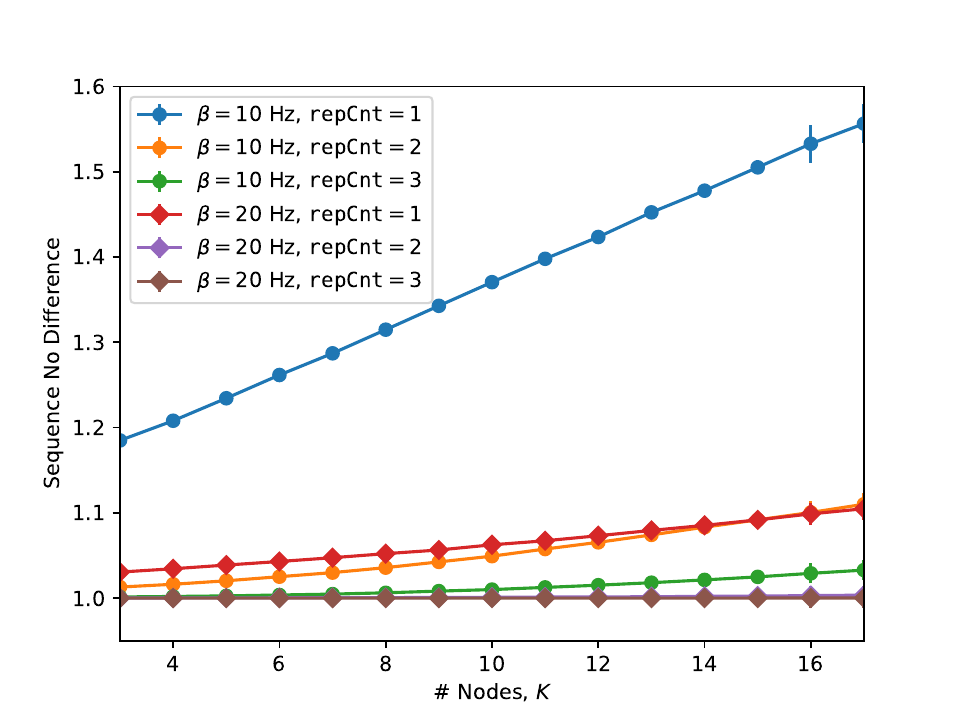}}
  \subfloat[$P=80\,\%$]{\includegraphics[width=0.5\textwidth]{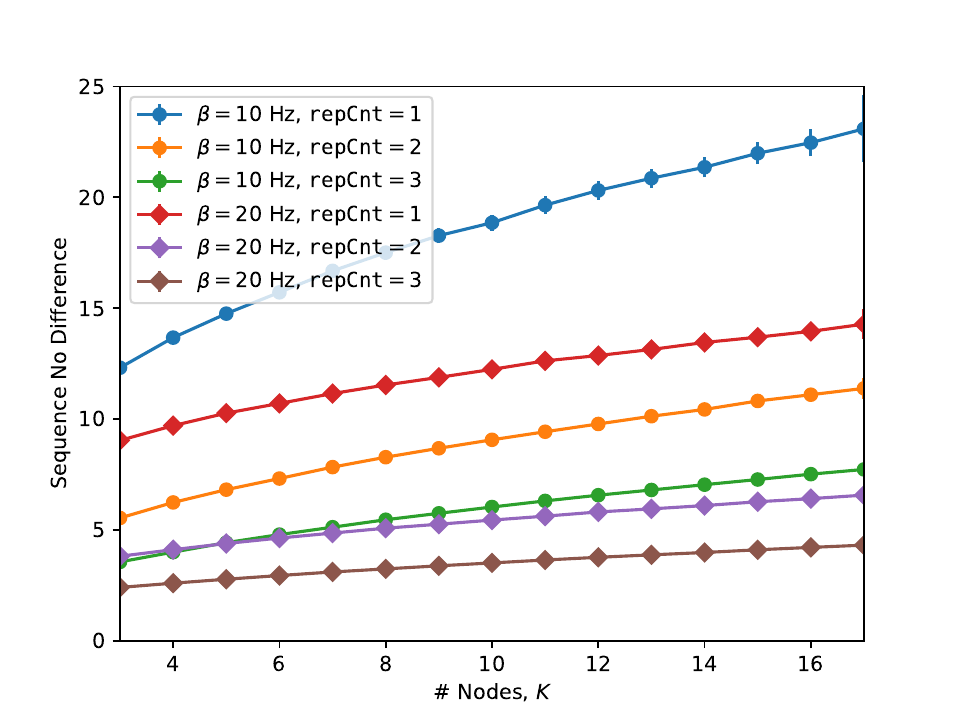}}
  \caption{Average sequence number gap for per-hop PER $P=50\,\%$ and
    $P=80\,\%$ with summaries, varying \texttt{repCnt} and beaconing
    rate.}
  \label{fig:experiment-one-avg-seqno-gap-50-80-percent-with-summaries}
\end{figure*}

In Figure \ref{fig:experiment-one-avg-seqno-gap-20-percent} we show
the average sequence number gap for a per-link packet error rate of
$P=20\,\%$, both with and without summarization, and for varying
beaconing rates and \texttt{repCnt} values. In Figure
\ref{fig:experiment-one-avg-seqno-gap-50-80-percent-with-summaries} we
show the average sequence number gap for $P=50\,\%$ and $P=80\,\%$, in
both cases with summarization being enabled, as the results without
summarization are very poor. For the lower per-link PER of $P=20\,\%$
the summarization mechanism achieves practically loss-free propagation
between producer and consumer with virtually no difference between the
different beaconing rates and \texttt{repCnt} values, even over the
maximum number of 16 hops. The results with summarization for the
higher per-link PER values of $P=50\,\%$ and $P=80\,\%$ (shown in
Figure
\ref{fig:experiment-one-avg-seqno-gap-50-80-percent-with-summaries})
show that the best-performing variants are consistently those with the
higher beaconing rate of 20\,Hz and at least two repetitions, followed
by the variant with 10\,Hz rate and three repetitions. This is likely
due to the fact that with a higher beaconing rate the average time for
an update to make the next hop (either directly or through the
update-request mechanism) becomes smaller, and therefore the
probability that an update reaches the consumer node within the
allotted five seconds of lifetime becomes higher --- if an update takes
longer than five seconds to propagate, it might be ``over-written'' by
the following update and never reach the consumer. It is also
interesting to note that even for a per-hop PER of 80\,\% and $K=17$
nodes the average gap size for 20 Hz beaconing rate and three
repetitions is below five. Furthermore, consistently the variants with
only one repetition behave the worst.

We now consider the case without summarization for $P=20$\,\% (compare
Figure \ref{fig:experiment-one-avg-seqno-gap-20-percent}).  The
average sequence number gap becomes very large for just one repetition
($\mathtt{repCnt}=1$), being around 405 for the 10 Hz beaconing rate
and 525 for the 20 Hz beaconing rate for $K=17$ nodes (likely a result
of higher beacon collision rates for $\beta=20$\,Hz, see below),
whereas adding a second and third repetition brings the gap size down
drastically (to around 4.8 for two repetitions and both beaconing
rates, and 1.6 for three repetitions and both beaconing rates). A
related finding emerges when comparing the simulation results against
a very simple and idealized probabilistic model of propagation without
summarization: with a per-link packet error rate of $P$, a number
$\mathtt{repCnt}$ of independent repetitions of an update per hop, and
$K-1$ hops, and furthermore assuming independent and statistically
identical links, the probability that an update actually reaches the
consumer can be expressed as
\begin{equation}
  \label{eq:experiment-one:reception-probability}
  Q=\left(1 - P^{\mathtt{repCnt}}\right)^{K-1},
\end{equation}
and the average sequence number gap in this model emerges as the
expectation $1/Q$ of a geometric random variable for the number of
trials required until (and including) a success. We find that the
expected sequence number gap from this model is noticeably smaller
than the average sequence number gap observed in our simulations. For
example, for $\mathtt{repCnt}=2$, $K=17$ and $P=0.2$ the expected
sequence number gap is $\approx 1.92$, while in our simulations we
have observed an average gap of around 4.8. This loss in reliability
is likely the result of beacon collisions (direct or hidden-terminal
collisions), which are known to be a significant issue for local
broadcasts (see Section~\ref{sec:related-work}), and which the simple
analytical model given in Equation
\eqref{eq:experiment-one:reception-probability} does not account for.

A key conclusion is that the summarization mechanism is indispensable
from a reliability perspective, so in the remaining paper we only
consider VarDis with summarization enabled.

\begin{figure*}
  \subfloat[With summaries]{\includegraphics[width=0.5\textwidth]{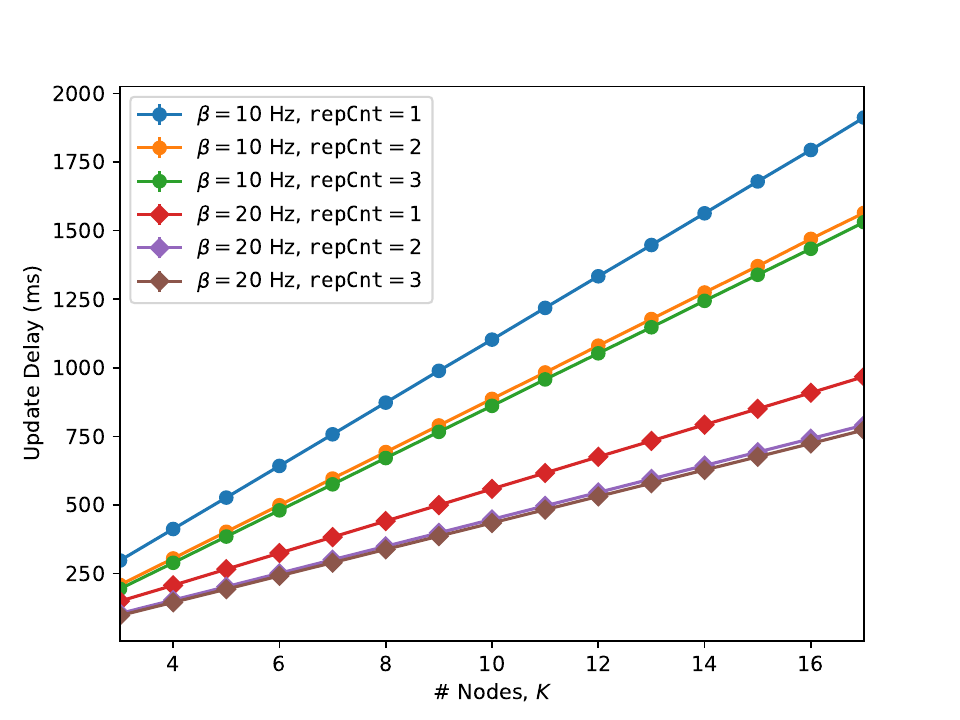}}
  \subfloat[Without summaries]{\includegraphics[width=0.5\textwidth]{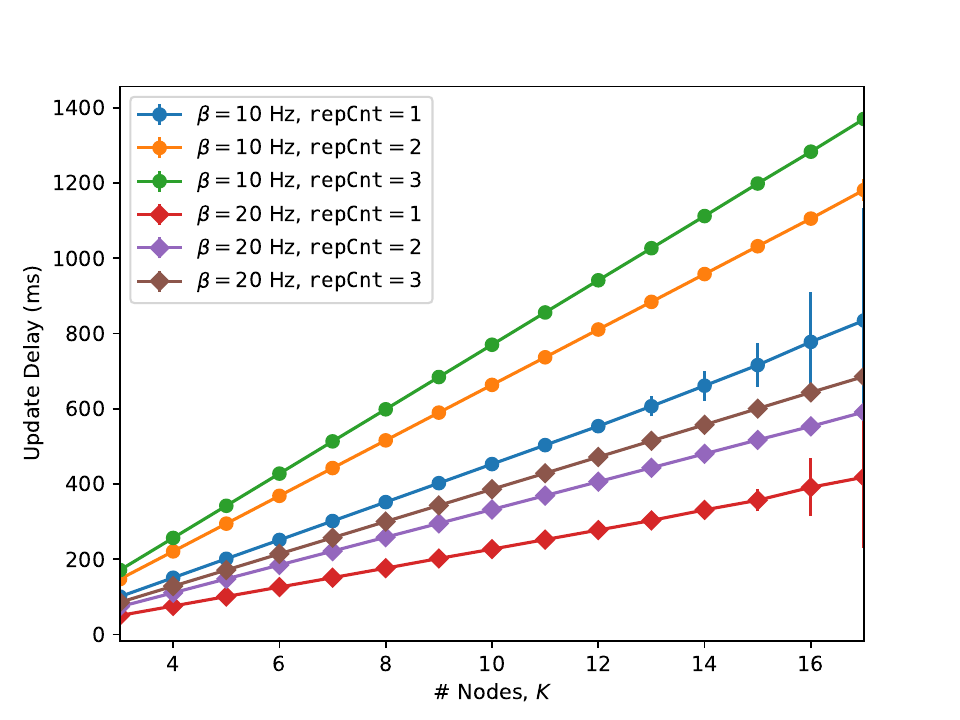}}
  \caption{Average update delay for per-hop PER $P=20\,\%$, with and
    without summaries, varying \texttt{repCnt} and beaconing rate.}
  \label{fig:experiment-one-avg-update-delay-20-percent}
\end{figure*}

\begin{figure*}
  \subfloat[$P=50\,\%$]{\includegraphics[width=0.5\textwidth]{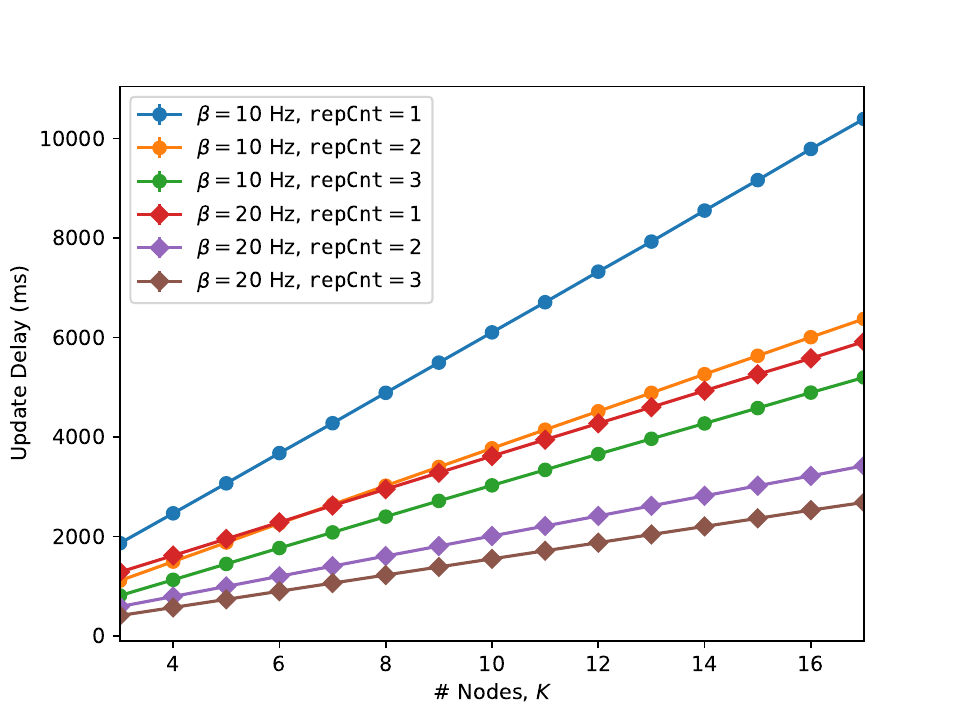}}
  \subfloat[$P=80\,\%$]{\includegraphics[width=0.5\textwidth]{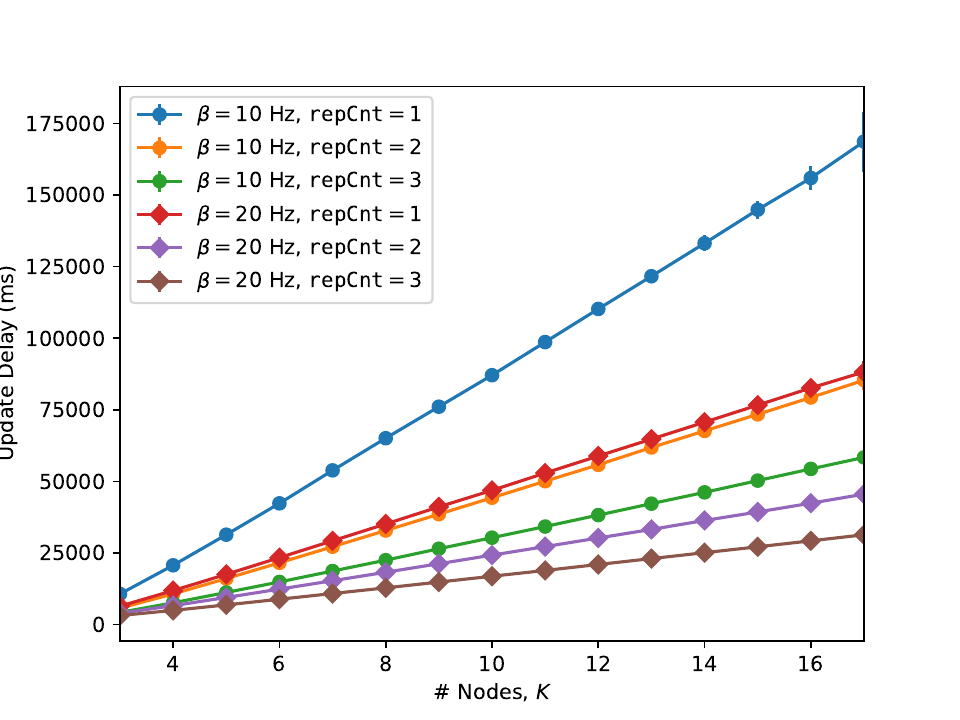}}
  \caption{Average update delay for per-hop PER $P=50\,\%$ and
    $P=80\,\%$ with summaries, varying \texttt{repCnt} and beaconing
    rate.}
  \label{fig:experiment-one-avg-update-delay-50-80-percent-with-summaries}
\end{figure*}

In Figure \ref{fig:experiment-one-avg-update-delay-20-percent} we show
the average update delay (travel time between producer and consumer
for a successful update) for a per-hop PER of 20\,\%, with and without
summarization and for varying beacon rate and \texttt{repCnt}.  In
Figure
\ref{fig:experiment-one-avg-update-delay-50-80-percent-with-summaries}
we show the average update delay for per-hop PER values of 50\,\% and
80\,\%, using summaries, again for varying beacon rate and
\texttt{repCnt}. Unsurprisingly, in all cases the average update delay
increases linearly with the number of hops, and the update delay for
20\,Hz beaconing rate is overall smaller than for their 10\,Hz
counterparts. Furthermore, for the same parameters, the delay results
without summaries will always be at least as good as for the case with
summaries --- recall that for the update delay we only consider those
updates that actually reach the destination, and these tend to travel
faster when the summary and update-request mechanism are not used.

When summarization is enabled, in all cases the update delay for
20\,Hz and three repetitions is the smallest, followed by 20\,Hz and
two repetitions. This suggests that receiving updates via repetitions
is faster than having to request them via the request-update
mechanism. Note that for $P=50\,\%$ and $P=80\,\%$ the update delay
for 10\,Hz beaconing rate and three repetitions is better than for
20\,Hz and one repetition. A likely explanation is that for these high
per-link PER values a node is more likely to have to resort to the
summary and update-request mechanism, compared to the case where just
one repetition is being used.

\subsection{Sensitivity Analysis: Method}
\label{subsec:results:fixed-density:sensitivity-method}

We perform a sensitivity analysis on a fixed-density grid network
(compare Section~\ref{subsec:system-model:deployment-traffic}) to
investigate the relative importance of key BP and VarDis parameters.

We first very briefly summarize the methodology used for the
sensitivity analysis, then we present our findings in
Section~\ref{subsec:results:fixed-density:sensitivity-results}. The
method adopted here is the same as in
\cite{Moravejosharieh:Willig:15}, and our presentation of this method
in this paper follows the one given in
\cite{Moravejosharieh:Willig:15} closely.

We apply the response surface methodology (RSM)
\cite{Myers:Montgomery:02}, \cite{Jain:91}. In this approach, the
dependency of a response variable (here: the update delay or the
average gap size) on a number of input factors is investigated using a
regression model. Each factor takes two different values, interpreted
as minimum and maximum values.

We have adopted a second-order polynomial regression model. In
general, the response variable $Y$ under consideration (average delay
or average gap size) is assumed to depend on the $k$ factors $x_i$ as
follows:
\begin{equation}
  \label{eq:regression-ansatz}
  Y = \alpha_0 + \sum_{i=1}^k \alpha_i x_i
     + \sum_{i=1}^k \sum_{j<i} \alpha_{i,j} x_i  x_j
\end{equation}
For each combination of factor values
$\vect{x} = (x_1,\ldots,x_k) \in \set{-1,1}^k$ we simulate to obtain a
response value $y_{x_1,\ldots,x_k}$, then we determine the parameters
$\alpha_\nu$ to minimize the least-squares error between the
regression model and the observed responses. The coefficients
$\alpha_0$, $\alpha_i$ and $\alpha_{i,j}$ are known as the intercept
(it represents the average value of all responses observed for the
various parameter combinations), linear, and mixed coefficients (or
interactions), respectively. It is common to encode the minimum and
maximum values for each factor not directly, but rather as $-1$ and
$1$, so all factors enter with the same magnitude. Units are
ignored. After computing the regression coefficients $\alpha_\nu$, we
determine the \emph{sum-of-squares-total} (SST)
\begin{equation}
  \label{eq:sst}
  SST = \sum_{\vect{x} \in \set{-1,1}^k} (y_x - \bar{y})^2
\end{equation}
representing the total amount of variation between the simulated
response values and the average response $\bar{y}$ (taken over all
parameter combinations), and the \emph{sum-of-squares-errors}
\begin{eqnarray}
  \label{eq:sse}
  \lefteqn{SSE = } \\
  \nonumber & & \sum_{\vect{x} \in \set{-1,1}^k} \left(y_x - \left(
          \alpha_0 + \sum_{i=1}^k \alpha_i x_i
     + \sum_{i=1}^k \sum_{j<i} \alpha_{i,j}  x_i  x_j\right)\right)^2
\end{eqnarray}
representing the total error between regression model and observed
responses. Exploiting that
\begin{equation}
  SST = 2^k (\alpha_1^2 + \ldots + \alpha_k^2 + \alpha_{2,1}^2
  + \alpha_{3,1}^2
  + \alpha_{3,2}^2 +
  \ldots + \alpha_{k,k-1}^2)
\end{equation}
holds, the relative impact of factor $k$ or any interaction $k,j$ is
given as:
\begin{equation}
  \label{eq:impact-of-factors}
  \frac{2^k \alpha_k^2}{SST}  \qquad , \qquad \frac{2^k \alpha_{k,j}^2}{SST}
\end{equation}
These values express the contribution of individual factors with
respect to the total observed variation.

A key quantity to assess the fit of the regression model is the
coefficient of determination or $R^2$ value, given by:
\begin{equation}
  \label{eq:coefficient-of-determination}
  R^2 = \frac{SST-SSE}{SST}
\end{equation}
where higher values are better. This quantity is a relative measure of
the total variation that can be explained by the regression model.

\subsection{Fixed-Density Networks: Sensitivity Analysis Results}
\label{subsec:results:fixed-density:sensitivity-results}

We perform this experiment using fixed-density $K \times K$ grid
networks with horizontal and vertical links of length $L_P$ having
packet error rates of $P\,\%$ (see Section
\ref{subsec:system-model:deployment-traffic}, compare also
Figure~\ref{fig:grid-deployment}). Data can only move along the
horizontal and vertical links, as the diagonal links are long enough
to have a PER of nearly 100\%{}.  Each of the $K^2$ drones in the
network produces a variable with update inter-arrival times drawn from
an exponential distribution with an average inter-arrival time of
$\lambda$, which is varied. We have used the following three factors:
\begin{itemize}
\item beaconing rate $\beta\in\set{10\,\mbox{Hz}, 20\,\mbox{Hz}}$, 
\item repetition count $\mathtt{repCnt}\in\set{1,3}$, and
\item maximum number of summaries $\mathtt{maxSumCnt}\in\set{10, 20}$.
\end{itemize}
We have varied all these factors in twelve different settings: the
packet error rates $P$ along vertical / horizontal links are chosen
from $P\in\set{10\,\%, 20\,\%}$ (with the distances $L_P$ being set
accordingly), the grid size parameter $K$ is chosen from
$K\in\set{9,13}$, and for the average update inter-arrival time
$\lambda$ we have used
$\lambda\in\set{300\,\mbox{ms}, 1000\,\mbox{ms},
  3000\,\mbox{ms}}$. For each of these twelve settings we have
simulated all $2^3=8$ combinations of the above factor values. For
these simulations we use a larger maximum beacon size of 300\,B to
account for the larger number of nodes.

\begin{table*}
  \begin{center}
    \begin{tiny}
      \begin{tabular}{|l|l|l|l|l|l|l|l|l|l|}
        \hline
        \textbf{$K$} & \textbf{Link PER (\%)} & \textbf{Upd.\ period (ms)} & $\mathbf{R^2}$ (\%) & \textbf{Average} & \textbf{Min.} & \textbf{Max.} & \textbf{\% Contr.\ \texttt{repCnt}} & \textbf{\% Contr.\ $\beta$} & \textbf{\% Contr.\ Interactions}  \\ 
        \hline
        9 & 10 & 300 & 100.0 & 2345.71 & 521.03 & 5222.15 & 64.85 & 26.17 & 8.96 \\ 
        9 & 10 & 1000 & 100.0 & 1156.78 & 388.2 & 3013.43 & 26.71 & 47.68 & 25.61 \\ 
        9 & 10 & 2000 & 99.99 & 545.79 & 341.97 & 767.84 &  & 98.41 & 1.22 \\ 
        \hline
        9 & 20 & 300 & 100.0 & 2702.51 & 1283.88 & 4923.47 & 34.86 & 55.3 & 9.75 \\ 
        9 & 20 & 1000 & 100.0 & 1635.7 & 678.19 & 2949.04 & 3.58 & 87.15 & 8.98 \\ 
        9 & 20 & 2000 & 99.91 & 962.63 & 462.72 & 1596.73 & 5.86 & 92.37 & 0 \\ 
        \hline
        13 & 10 & 300 & 100.0 & 6663.97 & 1305.74 & 13640.6 & 75.83 & 20.86 & 3.3 \\ 
        13 & 10 & 1000 & 100.0 & 5769.56 & 767.22 & 15619.07 & 57.21 & 25.64 & 17.14 \\ 
        13 & 10 & 2000 & 100.0 & 3169.2 & 615.35 & 10089.21 & 30.75 & 39.34 & 29.91 \\ 
        \hline
        13 & 20 & 300 & 100.0 & 7416.38 & 3422.17 & 13379.16 & 52.91 & 39.46 & 7.6 \\ 
        13 & 20 & 1000 & 100.0 & 6305.68 & 2624.26 & 13540.75 & 29.27 & 52.8 & 17.89 \\ 
        13 & 20 & 2000 & 100.0 & 3880.74 & 1163.29 & 8247.26 & 8.8 & 74.3 & 16.79 \\ 
        \hline
      \end{tabular} 
    \end{tiny}
  \end{center}
  \caption{Results of sensitivity analysis for update delay.}
  \label{tab:exp3:delay-sensitivity}
\end{table*}

The sensitivity analysis results for the average update delay are
presented in Table \ref{tab:exp3:delay-sensitivity}. We have included
the $R^2$ values, the average or intercept value, the minimum and
maximum observed values, the percentage contributions of the factors
\texttt{repCnt} and $\beta$, and the combined contributions of the
interactions between factors. We have not included contribution
results for \texttt{maxSumCnt}, since neither the percentage
contribution of its corresponding factor nor of any interaction
involving this factor exceed 1\,\%. Stated differently: the
\texttt{maxSumCnt} parameter has very limited impact.  Furthermore, in
the table we have also omitted entries for contributions below the
1\,\% threshold. The following points are noteworthy:
\begin{itemize}
\item The chosen regression model (Equation
  \eqref{eq:regression-ansatz}) represents the data very well, as all
  $R^2$ values are (very close to) 100\,\%.
\item Generally, as the update period $\lambda$ increases, the
  contribution of \texttt{repCnt} decreases and the contribution of
  $\beta$ becomes (often vastly) dominant.
\item The average delay values decrease with increasing update
  period. We attribute this to a reduced update load, in particular to
  the resulting ability of nodes to include a higher fraction of
  updates currently scheduled for transmission into the next beacon,
  therefore reducing their waiting times.
\item When investigating which values for the factors achieve the
  minimum delay observations in the twelve settings, we find that in
  all cases the beaconing rate has to be 20\,Hz and the value of
  \texttt{maxSumCnt} has to be 10 (though the difference to
  \texttt{maxSumCnt}=20 is very small). For the larger network
  ($K=13$), in all but one case the best value of \texttt{repCnt} is
  1, only for $P=20\,\%$ and $\lambda = 2\,\mbox{s}$ is it 3, but in
  that case the contribution of the \texttt{repCnt} factor is quite
  small ($\approx 8.8\,\%$). For the smaller network, we find that
  \texttt{repCnt}=1 gives minimal delay for $\lambda = 0.3\,\mbox{s}$,
  and for $\lambda=1\,\mbox{s}$ at $P=10\,\%$, in the other cases
  \texttt{repCnt}=3 is minimal. However, again in these other cases
  the contribution of the \texttt{repCnt} parameter is quite
  small. Hence, the overall recommendation (particularly for larger
  deployments) is to generally pick higher beaconing rates while
  keeping \texttt{repCnt} at its lowest value of 1.
\end{itemize}

\begin{table*}
  \begin{center}
    \begin{tiny}
      \begin{tabular}{|l|l|l|l|l|l|l|l|l|l|}
        \hline
        \textbf{$K$} & \textbf{Link PER (\%)} & \textbf{Upd.\ period (ms)} & $\mathbf{R^2}$ (\%) & \textbf{Average} & \textbf{Min.} & \textbf{Max.} & \textbf{\% Contr.\ \texttt{repCnt}} & \textbf{\% Contr.\ $\beta$} & \textbf{\% Contr.\ Interactions}  \\ 
        \hline
        9 & 10 & 300 & 100.0 & 2.57 & 1.62 & 3.98 & 36.51 & 55.72 & 7.73 \\ 
        9 & 10 & 1000 & 100.0 & 1.2 & 1.08 & 1.41 & 10.46 & 72.15 & 17.34 \\ 
        9 & 10 & 2000 & 99.96 & 1.06 & 1.03 & 1.08 & 8.1 & 91.71 & 0 \\ 
        \hline
        9 & 20 & 300 & 99.93 & 4.33 & 2.78 & 6.31 & 9.14 & 89.3 & 0 \\ 
        9 & 20 & 1000 & 99.9 & 1.56 & 1.18 & 2.02 & 14.09 & 85.13 & 0 \\ 
        9 & 20 & 2000 & 100.0 & 1.16 & 1.05 & 1.31 & 31.99 & 63.65 & 4.3 \\ 
        \hline
        13 & 10 & 300 & 100.0 & 4.82 & 2.81 & 7.61 & 40.92 & 53.45 & 5.58 \\ 
        13 & 10 & 1000 & 100.0 & 1.72 & 1.23 & 2.55 & 36.6 & 50.11 & 13.21 \\ 
        13 & 10 & 2000 & 100.0 & 1.19 & 1.05 & 1.45 & 15.74 & 60.37 & 23.83 \\ 
        \hline
        13 & 20 & 300 & 99.82 & 8.84 & 5.41 & 14.51 & 27.68 & 65.11 & 5.38 \\ 
        13 & 20 & 1000 & 99.88 & 2.75 & 1.77 & 3.95 & 11.1 & 87.52 & 0 \\ 
        13 & 20 & 2000 & 99.9 & 1.56 & 1.14 & 2.05 & 15.78 & 83.43 & 0 \\ 
        \hline
      \end{tabular} 
    \end{tiny}
  \end{center}
  \caption{Results of sensitivity analysis for average gap size.}
  \label{tab:exp3:gapsize-sensitivity}
\end{table*}

The corresponding sensitivity analysis results for the average
sequence number gap size between the bottom-right corner node and the
top-left corner node are presented in Table
\ref{tab:exp3:gapsize-sensitivity}.  Again, information about the
\texttt{maxSumCnt} parameter has been omitted, as it never contributes
more than $1\,\%$ to the observed variation of results. The following
points are noteworthy:
\begin{itemize}
\item Again the chosen regression model explains the variation very
  well, with $R^2$ values very close to 100\,\%.
\item Generally, the larger the average update period $\lambda$, the
  smaller the average gap size, which again can be attributed to a
  reduced update load, faster forwarding of updates and therefore
  reduced chances of an update being over-written by the next update
  of the same variable while in transit.
\item The contribution of the beaconing rate $\beta$ is dominant or
  even very dominant over \texttt{repCnt} in all cases, with the
  dominance mostly being more pronounced for more error-prone links
  ($P=20\,\%$).
\item When investigating the factor values giving the minimum average
  gap size in the twelve settings, we find that in all cases the
  beacon rate is $\beta=20\,\mbox{Hz}$, the \texttt{maxSumCnt}
  parameter is always 20 (but recall that it is largely
  inconsequential), and the \texttt{repCnt} parameter is three in most
  settings, except for $K=13$, $P=10\,\%$ and
  $\beta\in\set{0.3\,\mbox{s}, 1\,\mbox{s}}$, and for $K=9$,
  $P=10\,\%$ and $\beta=0.3\,\mbox{s}$.
\end{itemize}

\begin{figure*}
    \foreach \k in {5, 9, 13} {
        \subfloat[$K=\k$]{\includegraphics[width=0.5\columnwidth]{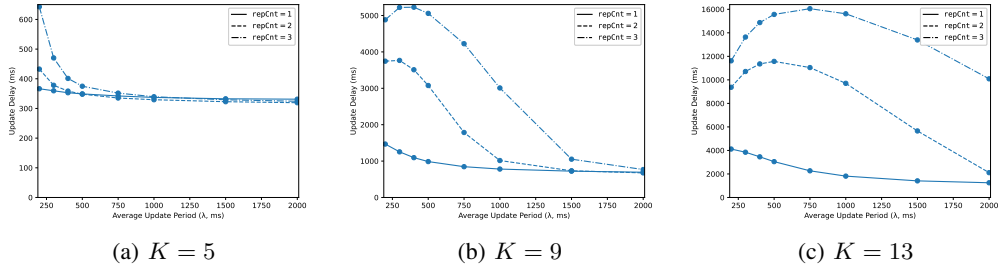}}
    }
    \centering

    \caption{Average update delay for varying \texttt{repCnt} with
        $\mathtt{maxSumCnt}=10, P=10\%$, and $\beta=10\,\mbox{Hz}$.}
    \label{fig:experiment-three-repcnt-effect-update-delay}
\end{figure*}

\begin{figure*}
    \foreach \k in {5, 9, 13} {
        \subfloat[$K=\k$]{\includegraphics[width=0.5\columnwidth]{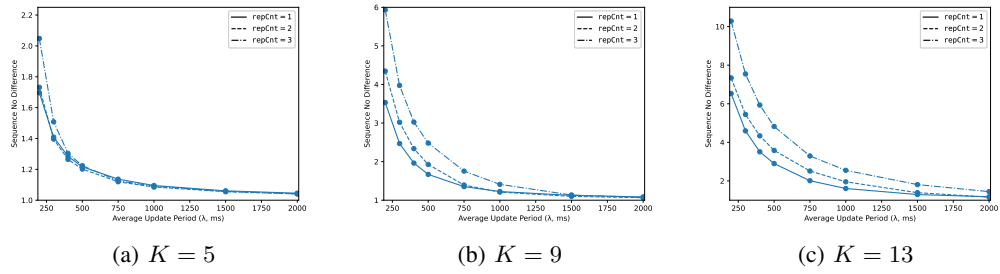}}
    }
    \centering

    \caption{Average sequence number difference for varying
        \texttt{repCnt} with $\mathtt{maxSumCnt}=10, P=10\%$,
        and $\beta=10\,\mbox{Hz}$.}
    \label{fig:experiment-three-repcnt-effect-seqno-diff}
\end{figure*}

Finally, to give an additional impression on how the results depend on
the average update period $\lambda$, we show in
Figures~\ref{fig:experiment-three-repcnt-effect-update-delay} and
\ref{fig:experiment-three-repcnt-effect-seqno-diff} the effect of
varying $\lambda$ for different values of \texttt{repCnt} and $K$ on
the update delay and sequence number difference. Note that we only
show results for the 10\,\mbox{Hz} beacon rate, as the 20\,\mbox{Hz}
values show the same trends.  In scenarios with an already high update
load (e.g., short average update periods or large networks),
increasing the \texttt{repCnt} value further reduces VarDis'
performance in terms of both the average sequence number difference
and the update delay. However, for small networks and longer average
update periods, $\mathtt{repCnt}=2$ performs slightly better on both
update delay and sequence number difference than both other options.

\subsection{Fixed-Density Networks: Comparison Against Flooding Protocol}
\label{subsec:results:fixed-density:flooding-comparison}

We compare the performance of VarDis to the flooding protocol
described in Section~\ref{subsec:system-model:flooding-protocol}, in
which each variable update triggers a flooding operation (with each
node repeating a packet \texttt{repCnt} times). A flooding protocol
like that can be seen as a natural competitor for VarDis. We again use
a fixed-density grid network. Based on the results of our sensitivity
analysis
(Section~\ref{subsec:results:fixed-density:sensitivity-results}), we
only compare against the $\beta=20\,\mbox{Hz}$ and
$\texttt{maxSumCnt}=10$ results.

We perform the comparison on a fixed-density grid network with
$K\in\set{5,7,9,11}$ nodes on each side of the grid (compare
Section~\ref{subsec:system-model:deployment-traffic}), and having
packet error rates $P\in\set{10\,\%, 20\,\%}$ on the horizontal and
vertical links (corresponding to distances
$L_P\in\set{255\,\mbox{m}, 263\,\mbox{m}}$). We furthermore considered
two different values for the update inter-arrival times, namely
$\lambda\in\set{200\,\mbox{ms}, 1000\,\mbox{ms}}$. We restrict the
comparison to a metric we have not used so far, namely the percentage
of variable updates issued on the chosen producer that are received by
the chosen consumer (see
Section~\ref{subsec:system-model:deployment-traffic}) -- note that for
both protocols any received update at the consumer is guaranteed to be
strictly new, i.e.\ it is not a duplicate of an earlier received
update. The reason for this choice of metric is that the flooding
protocol involves queues and we have found these queues to be unstable
(i.e.\ continuously growing over time) for most of the considered
scenarios, which prohibits the calculations of average delay or gap
size results.

\begin{figure*}
  \subfloat[Link PER $L_P=10\,\%$, Update period $\lambda=200\,\mbox{ms}$]{\includegraphics[width=0.5\textwidth]{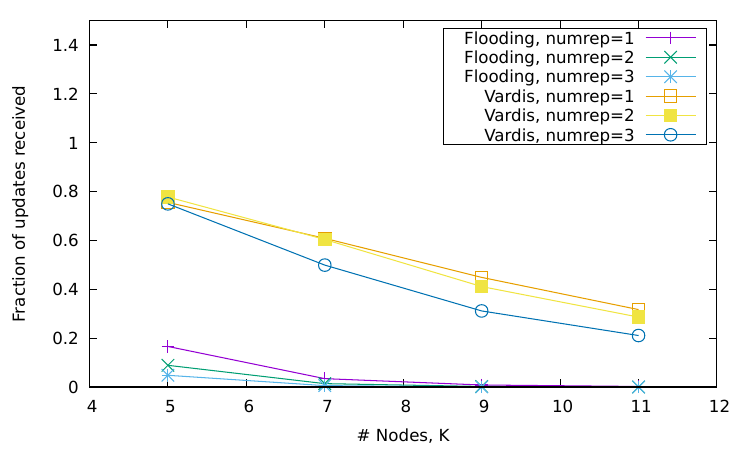}}
  \subfloat[Link PER $L_P=10\,\%$, Update period
  $\lambda=1000\,\mbox{ms}$]{\includegraphics[width=0.5\textwidth]{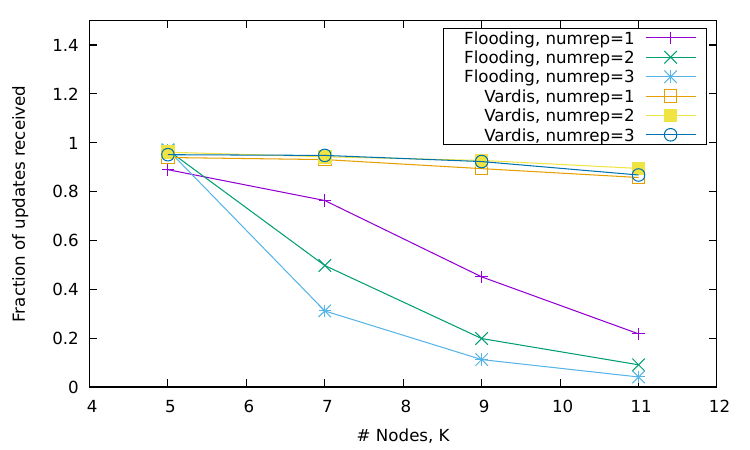}}
  \\
  \subfloat[Link PER $L_P=20\,\%$, Update period $\lambda=200\,\mbox{ms}$]{\includegraphics[width=0.5\textwidth]{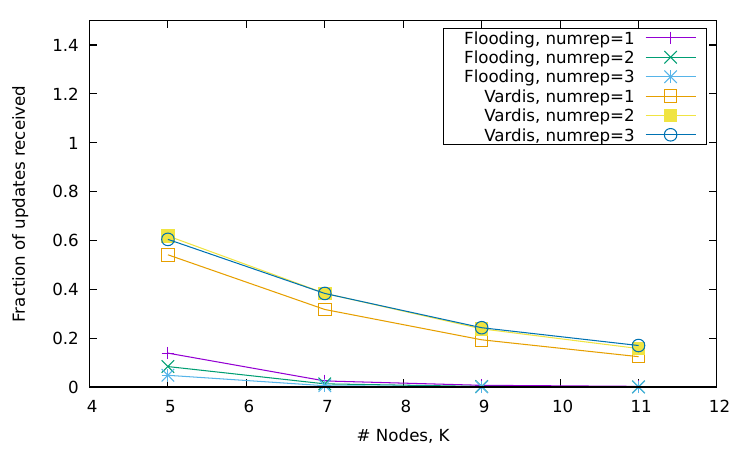}}
  \subfloat[Link PER $L_P=20\,\%$, Update period $\lambda=1000\,\mbox{ms}$]{\includegraphics[width=0.5\textwidth]{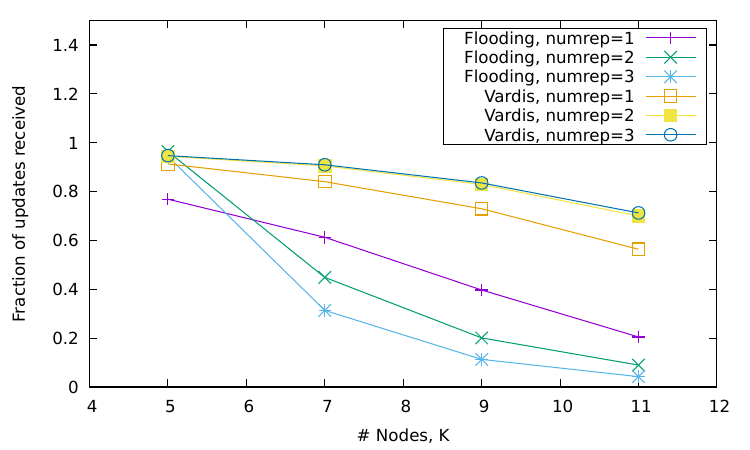}}
  \caption{Comparing percentage of received updates between the
    flooding protocol and VarDis}
  \label{fig:flooding-comparison:percentage-of-received-updates}
\end{figure*}

The results are shown in
Figure~\ref{fig:flooding-comparison:percentage-of-received-updates}. It
can be observed that VarDis consistently outperforms the flooding
protocol, except for the smallest network ($K=5$), where both
protocols perform on similar, nearly perfect levels. For both
protocols their performance decreases as the network size (and
therefore network load) increases. The difference can likely be
explained by VarDis remaining stable (the VarDis protocol does not
include any queues), where flooding does not. This is supported by
noting that for the flooding protocol the variants with
$\mathtt{repCnt}=1$ outperform those with
$\mathtt{repCnt}\in\set{2,3}$ for all but the smallest network size ---
note that higher $\mathtt{repCnt}$ values lead to higher packet
load and longer queues. When focusing on VarDis, it can be observed
that the variant with $\mathtt{repCnt}=2$ shows consistently optimal
or close-to-optimal performance, whereas the relative order of
$\mathtt{repCnt}=1$ and $\mathtt{repCnt}=3$ depends on the scenario,
with $\mathtt{repCnt}=1$ having the advantage for higher link packet
error rates.

\subsection{Variable-Density Networks: Comparison against Numerical
  Results and Effect of Inter-Beacon Times}
\label{subsec:results:variable-density:numerical-comparison-beaconing}

We again explore the impact of VarDis parameters on the reliability
and update delays, but now for a variable-density line network, where
drones may be able to reach more than one neighbour in each
direction.

We use the variable-density line network introduced in
Section~\ref{subsec:system-model:deployment-traffic}, and in
particular Figure~\ref{fig:line-deployment:variable-density}.  There
are $K$ drones in total, placed equidistantly on a line, with a fixed
distance of $D=1,120\,\mbox{m}$ between the first and last drones. The
first drone in the chain produces the chain's sole variable,
generating updates every 5\,s. The final drone in the chain is the
consumer for which we present average delay results. We do not show
results for the average sequence number gap, as it is consistently
extremely close to one for all considered parameter combinations.

As we have a fixed chain total length with only one producer, in our
simulations we have only varied three parameters: the number of drones
($K\in\set{6,7,\ldots,18}$), the VarDis repetition count
($\mathtt{repCnt}\in\set{1,2,3}$), and the beaconing rate
($\beta\in\set{10\,\mbox{Hz},20\,\mbox{Hz}}$).

\subsubsection{Line Network: Simulation Results}
\label{subsubsec:results:variable-density:numerical-comparison-beaconing:simulation-results}

\begin{figure*}
    \subfloat[$\beta=10\,\mbox{Hz}$]{\includegraphics[width=0.5\textwidth]{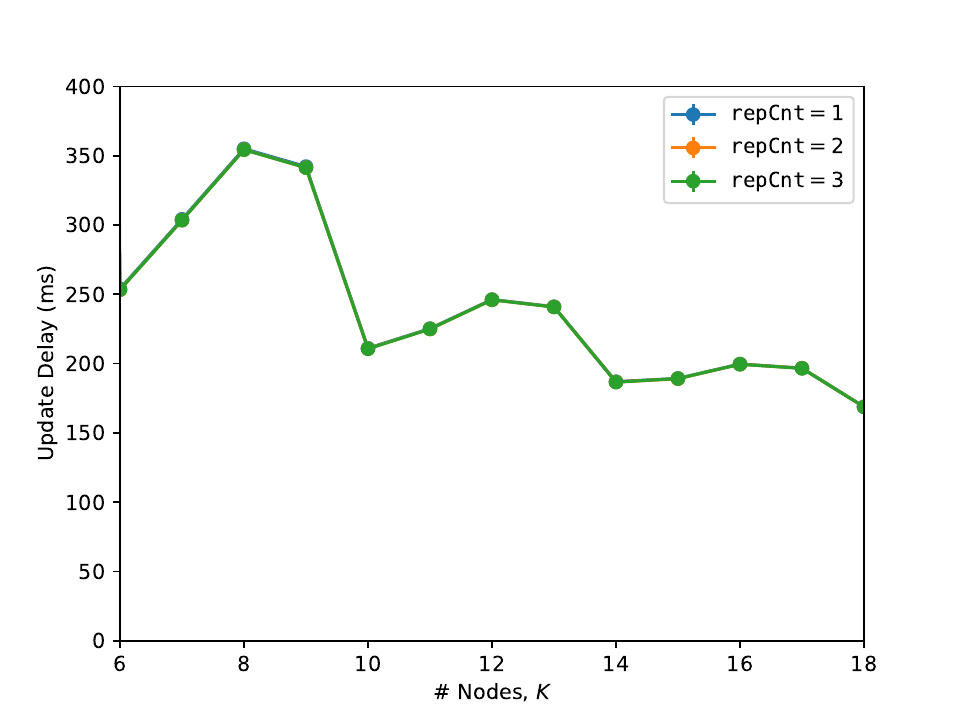}}
    \subfloat[$\beta=20\,\mbox{Hz}$]{\includegraphics[width=0.5\textwidth]{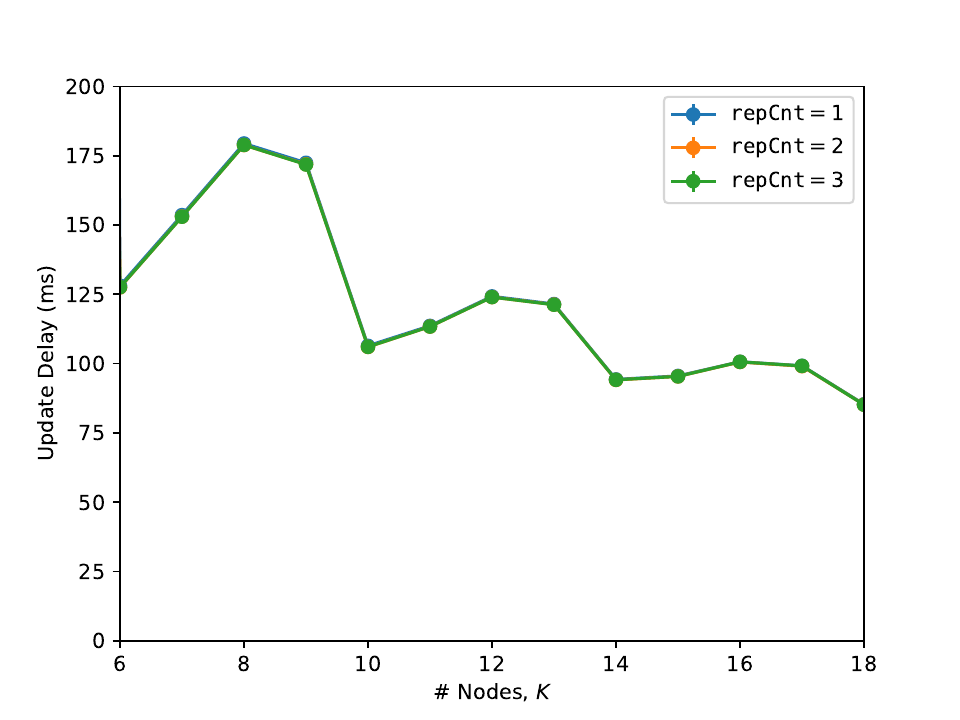}}
    \caption{Average update delay for varying \texttt{repCnt} and
             beaconing rate.}
    \label{fig:variable-density:line:-avg-update-delay}
\end{figure*}

In Figure~\ref{fig:variable-density:line:-avg-update-delay} we show
the average update delay. As the number of drones increases, the
update delay increases as well, before suddenly dropping to below the
initial update delay and then beginning to increase again. We
hypothesize that this is due to two inter-related effects. As $K$
increases, each node changes from having just one neighbour (a
neighbour being a node reachable in a single hop with reasonably high
packet success rate) to having two neighbours, then three neighbours
and so on. When a beacon transmitted by a node carrying a variable
update is received by $n$ other nodes further down the chain, and each
of these $n$ other nodes waits a random time $T_i$ before forwarding
the variable update further, then the time before \emph{any} of the
$n$ downward nodes forwards the beacon is
$\min\setof{T_i}{i=1,\ldots,n}$. In an example where each forwarder
transmits beacons without collisions and with an exponentially
distributed and independent inter-arrival time distribution of rate
parameter $\beta$, the average forwarding delay would become
$\frac{1}{n \beta}$. Furthermore, with reference to the specific
topology shown in Figure~\ref{fig:line-deployment:variable-density},
there is a possibility that this next forwarder is not the node
closest to the transmitter, but a node two or more hops closer to the
destination, so that the packet makes quicker geometric progress ---
this in fact happens with probability $\frac{n-1}{n}$. In
Figure~\ref{fig:variable-density:line:-avg-update-delay}, each
``bump'' corresponds to one further forwarder coming into reach of a
transmitting node.

The key conclusion is that VarDis benefits significantly from
increased node densities, speeding up delivery
substantially.

\subsubsection{Line Network: Numerical Results}
\label{subsubsec:results:variable-density:numerical-comparison-beaconing:numerical-results}

To at least partially validate our simulation model, we compare
simulation results against numerical results gained from a
time-homogeneous Discrete-Time Markov chain (DTMC) model of the
propagation of an update in a network, under a simplified protocol
model. From our DTMC model we can calculate the expected number of
``steps'' required until all nodes have received an update, with one
step corresponding to a single beacon transmission by a uniformly
chosen node. We then multiply this expected number of steps with the
expected time between steps to calculate the expected time until all
nodes have received an update.

We assume one single variable. For simplicity, we ignore beacon
collisions.  There are $K$ nodes in the network, numbered from $1$ to
$K$. One of the nodes is selected to be the producer of the variable,
e.g.\ node 1. The model allows for nodes to be arranged in a general
network, but in this study will only be evaluated for a line network.
Packet losses on the channel between nodes $i$ and $j$ happen
independently of packet losses on other channels with probability
$q_{i,j}$ --- we assume that $q_{i,i}=0$ for all nodes $i$. To
determine the $q_{i,j}$ for the variable-density line network
investigated in
Section~\ref{subsubsec:results:variable-density:numerical-comparison-beaconing:simulation-results},
we have used a step-wise approximation of the PER-versus-distance
function obtained from our simulator. We assume that beacons with and
without a variable update have the same length and therefore the same
packet loss probability. The model captures broadcasting effects,
where a single transmission can reach multiple receivers. For each
slot one node is randomly selected (with uniform distribution) to be
the beacon transmitter, independently of whether the node has the
update or not. Once a node has received the update, it will include it
in all subsequent beacons (and not just in \texttt{repCnt}
beacons). With this simplification, the state space of the model is
$\mathcal{S} = \set{0,1}^K \setminus \set{(0,\ldots,0)}$, where a
particular state $\vect{s}=(s_1,\ldots,s_K)$ records for each of the
nodes $i$ whether it actually has received the variable update
($s_i=1$) or not ($s_i=0$). Note that we have excluded the
uninteresting state $(0,0,\ldots,0)$ in which no node has the
update. To write down the state transition probabilities of moving
from state $\vect{s}\in\mathcal{S}$ to state $\vect{t}\in\mathcal{S}$
from one slot to the next (after one beacon transmission), the
following notations are convenient:
\begin{itemize}
\item Let $\vect{s}=(s_1,\ldots,s_K)\in\mathcal{S}$ and
  $\vect{t}=(t_1,\ldots,t_K)\in\mathcal{S}$ be the source and target
  state, such that $\vect{s} \le \vect{t}$ (i.e.\ for all
  $i\in\set{1,\ldots,K}$ it holds that $s_i \le t_i$). If the
  condition $\vect{s} \le \vect{t}$ is not fulfilled, then there must
  be one index $i$ such that $s_i=1$ and $t_i=0$ --- but such a
  transition will be impossible, as nodes never drop the
  update. Impossible transitions are assigned zero probability.
\item
  $\mathcal{R}(\vect{s},\vect{t})=\setof{i\in\set{1,\ldots,K}}{s_i=0
    \wedge t_i=1}$ is the set of nodes which did not have the update
  before the transmission but receive it through the transmission.
\item
  $\mathcal{N}(\vect{s},\vect{t})=\setof{i\in\set{1,\ldots,K}}{s_i=0
    \wedge t_i=0}$ is the set of nodes which did not have the update
  before the transmission and do not have it after the transmission.
\item $\mathcal{T}(\vect{s})=\setof{i\in\set{1,\ldots,K}}{s_i=1}$ is
  the set of nodes that already have the update at the start of the
  slot, and which are the desirable transmitters to help further
  dissemination.
\end{itemize}
With these notations, the state transition probabilities are shown in
Figure~\ref{fig:eq:transition-probabilities}.
\begin{figure*}
\begin{displaymath}
  p_{\vect{s},\vect{t}} =  \left\{\begin{array}{r@{\quad:\quad}l}
                                    0 & \vect{s} \nleq \vect{t} \\
                                    \frac{K-|\mathcal{T}(\vect{s})|}{K} + 
                                    \sum_{i\in\mathcal{T}(\vect{s})}
                                    \left( \frac{1}{K}
                                    \cdot \prod_{k\in\mathcal{N}(\vect{s},\vect{t})} q_{i,k}
                                    \right) & \vect{s} = \vect{t} \\
                                    \sum_{i\in\mathcal{T}(\vect{s})}
                                    \frac{1}{K}
                                    \left( 
                                    \cdot \prod_{j\in\mathcal{R}(\vect{s},\vect{t})} (1-q_{i,j})
                                    \right) \cdot \left(
                                    \prod_{k\in\mathcal{N}(\vect{s},\vect{t})} q_{i,k}
                                    \right) & \vect{s} \ne \vect{t},
                                              \vect{s} \le \vect{t} \\
                                  \end{array}
                                \right.
                              \end{displaymath}
                              \caption{Transition probabilities}
                              \label{fig:eq:transition-probabilities}
\end{figure*}
The first case is the case of an invalid state transition
($\vect{s} \nleq \vect{t}$), to which we assign probability zero. In
the third case, we consider a valid transition from $\vect{s}$ to
$\vect{t}\ne\vect{s}$, in which the set
$\mathcal{R}(\vect{s},\vect{t})$ of nodes newly receiving the beacon
transmission is non-empty. To achieve this transition, it is required
that one of the nodes currently having the beacon is chosen, i.e.\ a
node from $\mathcal{T}(\vect{s})$. Each member of this set is chosen
with probability $\frac{1}{K}$. Assume that node
$i\in\mathcal{T}(\vect{s})$ is chosen as sender. Node $i$'s
transmission then must be successfully received by all $j$ in the set
$\mathcal{R}(\vect{s},\vect{t})$ of receivers (which happens
independently with probability $1-q_{i,j}$ for receiver $j$), and it
must be missed by any node $k$ in the set of non-receivers
$\mathcal{N}(\vect{s},\vect{t})$, which happens independently with
probability $q_{i,k}$. As the channels are assumed to be independent,
the reception and non-reception events at the different nodes are
independent as well. The second case (where no node receives a new
update) is similar, but includes the option of choosing a beacon
transmitter that does not yet have the update, which happens with
probability $\frac{K-|\mathcal{T}(\vect{s})|}{K}$, and also is
simplified by noting that $\mathcal{R}(\vect{s},\vect{t})=\emptyset$
when $\vect{s}=\vect{t}$.

This Markov chain has one absorbing state: the state $(1,1,\ldots,1)$,
in which every node has the update. The desired average number of
steps for an update to reach \emph{all} nodes in the network can be
calculated as the expected number of steps required to reach state
$(1,1,\ldots,1)$, which in the literature is referred to as the
average hitting time for the target set
$\mathcal{A}=\set{(1,1,\ldots,1)}$ \cite{Norris:97}. In general, when
state $\vect{s}$ is the starting state of the chain, the average
number of steps $k_{\vect{s}}$ required to reach the target set
$\mathcal{A}\subset\mathcal{S}$ is given as the solution of the
following linear system (see \cite[Sec.\ 1.3]{Norris:97}):
\begin{equation}
  \label{eq:markov-model-hitting-times}
  k_{\vect{s}} =  \left\{\begin{array}{r@{\quad:\quad}l}
                           0 & \vect{s} \in \mathcal{A} \\
                           1 + \sum_{\vect{t}\notin\mathcal{A}}
                           p_{\vect{s},\vect{t}}\cdot k_{\vect{t}} &
                                                                     \vect{s} \notin\mathcal{A}
                         \end{array}
                       \right.
\end{equation}
For our given starting state $\vect{s_0}=(1,0,\ldots,0)$ (a state with
just one node at the end of the line having the update), we use the
solution of Equation~\eqref{eq:markov-model-hitting-times} to
numerically determine the average number $k_{\vect{s_0}}$ of steps ---
beacon transmissions --- required.\footnote{It should be noted that
  the reach of our model is quite limited, as the number of states for
  $K$ nodes is $2^K$, and the numerical solution of the system given
  by Equation~\eqref{eq:markov-model-hitting-times} hits singularities
  for $K>15$ when calculating with double precision.}

\begin{figure}
    \includegraphics[width=0.5\textwidth]{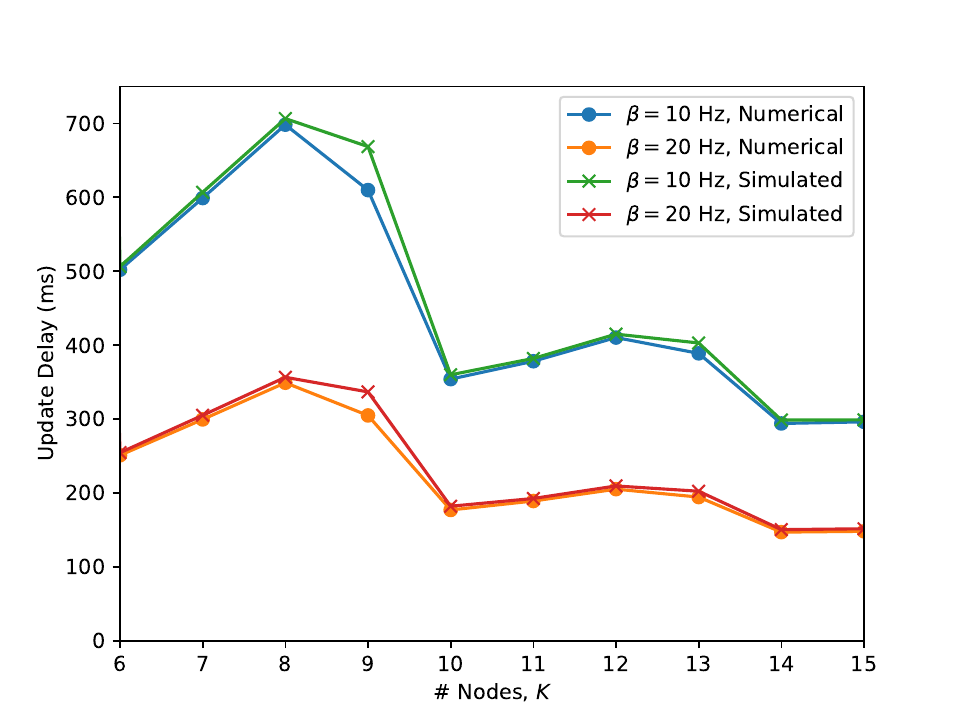}
    \caption{Average update delay for simulation and numerical model,
      using exponentially distributed beacon inter-arrival times}
    \label{fig:variable-density:line:avg-update-delay-numerical-comparison}
\end{figure}

For the comparison of numerical and simulation results, we consider
the case of exponentially distributed (with rate parameter $\beta$)
beacon inter-transmission times at the individual nodes. The beacon
arrivals at one node thus form a Poisson process of rate $\beta$
\cite{Karlin:Taylor:75}, and the overall beacon arrival process then
is the superposition of these Poisson processes. From the well-known
properties of Poisson processes, this superposition is again a Poisson
process of arrival rate $K\lambda$, and beacons indeed arrive
independently at the nodes with a uniform distribution. The numerical
and simulation results with exponentially distributed beacon
inter-arrival times are shown in Figure
\ref{fig:variable-density:line:avg-update-delay-numerical-comparison}. It
can be observed that simulation and numerical results show excellent
agreement, which confirms our trust into the simulation results.

\begin{figure}
    \includegraphics[width=0.5\textwidth]{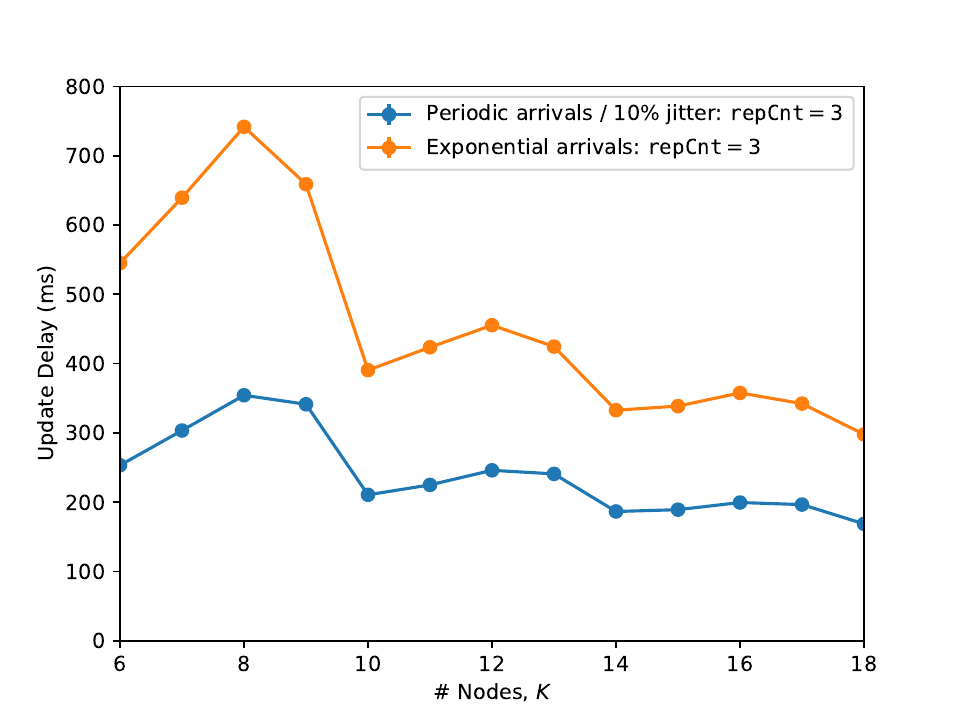}
    \caption{Average update delay for exponentially distributed
      interarrival times and periodic interarrival times with 10\,\%
      jiiter, for $\beta=10$ (simulation).}
    \label{fig:variable-density:line:avg-update-delay-iat-comparison}
\end{figure}

We finally consider the impact of the choice of the distribution of
inter-beacon generation times at the nodes. In Figure
\ref{fig:variable-density:line:avg-update-delay-iat-comparison} we
show simulation results for the delay, for varying $K$,
$\mathtt{repCnt}=3$ and $\beta=10$, for exponentially distributed
inter-arrival times with rate parameter $\beta$, and for periodic
inter-arrivals of average period $1/\beta$ and 10\,\% jitter. It can
be seen that the choice of the inter-beacon generation time
distribution makes a substantial difference. We illustrate the reason
for the difference for the case of $K=6$ and $\beta=10$\,Hz. For this
value of $K$, a node in our line network can practically only reach
the next node in the line, but not others, and the update hence has to
take 5 hops. First consider the case where the beacon inter-arrival
times are exponentially distributed with parameter $\beta$. As
mentioned above, the beacon arrivals at one node form an independent
Poisson process of rate $\beta$. Now suppose that an update arrives at
node 1. From the memoryless property of the exponential distribution,
the waiting time until the \emph{next} beacon transmission of node 1
is again exponentially distributed with parameter $\beta$. Hence, with
$\beta=10$\,Hz it takes on average 0.1\,s before the beacon
transmission takes place and node 2 receives the update. By the same
argument, node 2 has to wait on average 0.1\,s before its own next
beacon transmission, and so on. This means that with exponential
inter-arrival times the beacon will need 0.5\,s on average to reach
the final node in five hops distance to the producer. In the second
case we consider that at each node beacons arrive periodically with
10\,\% random jitter (i.e.\ the beacon inter-arrival times are drawn
uniformly from
$U\left[\frac{0.9}{\beta},\frac{1.1}{\beta}\right]$). The time between
update arrivals at the producer is comparatively long with five
seconds, and after this time (corresponding to 50 beacons on average),
the beacon transmissions between different nodes have become
unsynchronized. As a result, once any node $i$ other than the last one
has received the update from its predecessor, it will have to wait on
average half a period (i.e.\ $\frac{1}{2\beta}$\,s) before $i$
transmits the update to the next node. This means for $\beta=10$\,Hz
that the update takes on average 0.05\,s per hop, for a total average
time of 0.25\,s, twice as fast as in the case of exponential
inter-arrival times.

\subsection{Variable-Density Networks: The Grid Case and Network Capacity}
\label{subsec:results:variable-density:grid-case}

This experiment focuses on exploring the impact of drone density on
the performance of VarDis, but now in larger variable-density grid
networks, and focusing more on capacity questions, i.e.\ asking what
update load a network can carry subject to certain delay or gap size
constraints. The grid network has $K\times{}K$ nodes with a fixed
total sidelength $D=1,120\,\mbox{m}$ (cf.\
Section~\ref{subsec:system-model:deployment-traffic}). Each node
produces its own variable and generates updates with iid exponentially
distributed inter-update times, with an average inter-update time of
$\lambda$ seconds. Both $K$ and $\lambda$ are varied.

We define two different metrics to measure the VarDis protocols'
performance. The \emph{reliability update capacity} is the smallest
per-node average update period $\lambda$ (or the highest average
per-node update rate $\frac{1}{\lambda}$) at which the average
sequence number gap is still less than 1.5, whereas the \emph{delay
  update capacity} is the smallest per-node update period $\lambda$ at
which the average update delays is less than 250\,ms. Recall that
every node in the $K\times K$ network produces a variable, but the
delay and gap size are measured at the reference consumer for the
variable produced by the reference producer (compare
Section~\ref{subsec:system-model:deployment-traffic} and
Figure~\ref{fig:grid-deployment}), representing a ``worst-case''
scenario. The two `capacity' metrics introduced here characterize the
maximum amount of aggregated variable update traffic that the network
can carry subject to (somewhat arbitrary) delay and gap size
constraints.

Motivated by our findings in the sensitivity analysis
(Section~\ref{subsec:results:fixed-density:sensitivity-results},
obtained for a fixed-density network), we have fixed the beaconing
period to $\beta=20\,\mbox{Hz}$, and \texttt{maxRepCnt} to 1. We also
selected \texttt{maxSumCnt} to be 10 and used the larger maximum
beacon size of 300\,B. We simulated all networks for
$\lambda\in\set{0.15\,\mbox{s}, 0.175\,\mbox{s}, \ldots,
  0.475\,\mbox{s}, 0.5\,\mbox{s}, 0.75\,\mbox{s}, 1.0\,\mbox{s},
  1.5\,\mbox{s}, 2.0\,\mbox{s}}$.

\begin{figure*}
    \subfloat[Update Delay
    Capacity]{\includegraphics[width=0.5\textwidth]{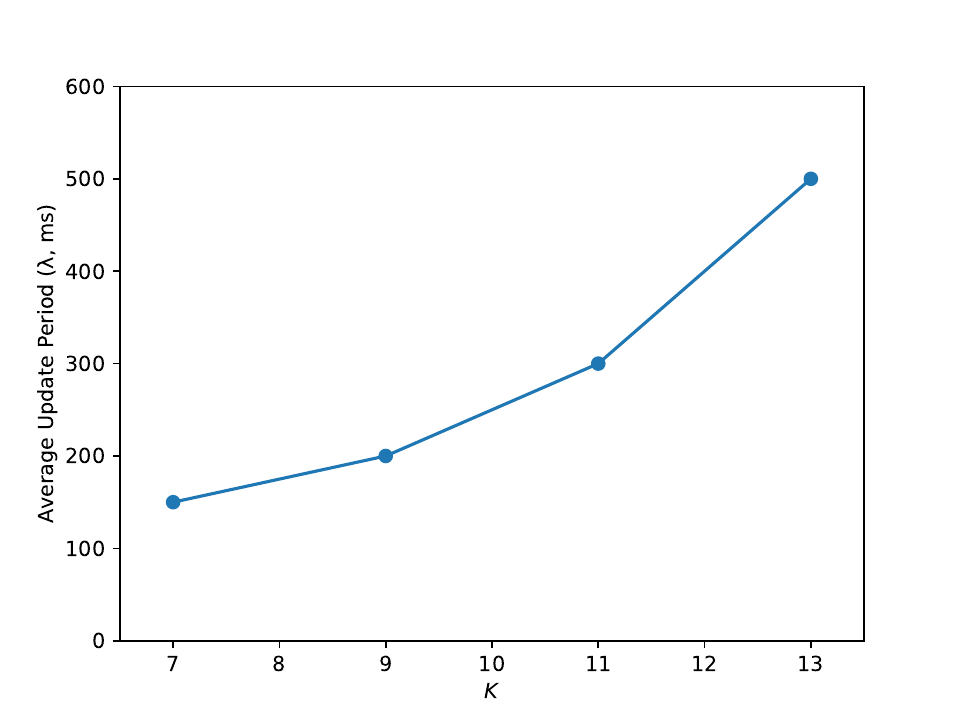}}
    \subfloat[Reliability Update
    Capacity]{\includegraphics[width=0.5\textwidth]{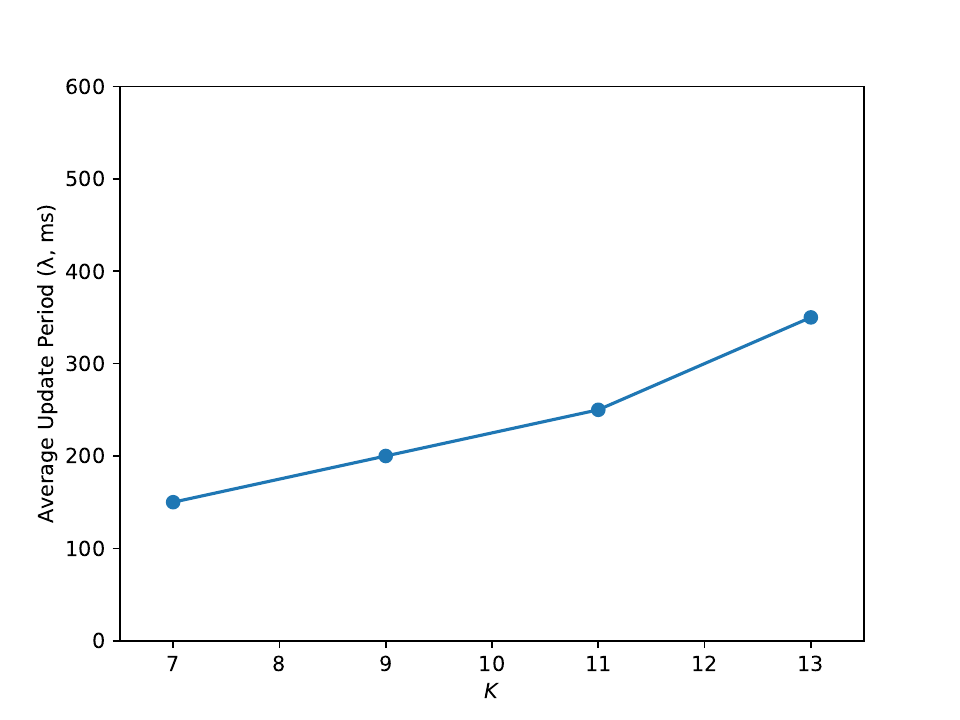}}
    \caption{The capacity of the VarDis protocol as the network density increases.}
    \label{fig:results:capacities}
\end{figure*}

The reliability and delay update capacity results are shown in
Figure~\ref{fig:results:capacities}. As the network size increases,
both the update and reliability capacity decrease as expected (the
minimal average periods $\lambda$ that can be supported become
larger), resulting from an increased network load due to having more
producers. However, it is quite remarkable that in a network with
$13\times 13=169$ nodes it is still possible for each node to send an
update approximately every 500\,ms before the gap size exceeds 1.5, or
for every node to send updates with a period of
$\approx 350\,\mbox{ms}$ before the delay between two nodes on
opposite corners exceeds 250\,ms. These results confirm that VarDis
can provide quite satisfying delay and reliability performance in
larger and denser networks.

\section{Conclusions}
\label{sec:conclusions}

In this paper we have presented the DCP / VarDis approach to global
data dissemination in networks of closely collaborating drones, for
example in a swarm or formation. VarDis offers the abstraction of a
variable, which is replicated into the entire network. It operates by
piggybacking modifying operations on a variable onto regularly
transmitted beacons, which for reasons of collision avoidance between
drones are required in many settings. We have analysed the performance
of VarDis in a range of conditions, and our results show that VarDis
provides a simple solution that can operate on top of a variety of
underlying wireless technologies and medium access control protocols.

There is substantial potential for future research. For example, the
current version of the VarDis summary mechanism is simple and
straightforward and can be optimized, e.g.\ by reducing the rate at
which summaries are transmitted for variables that have not changed in
a long time, or by aggregating summaries further. Another option is to
not re-broadcast a variable update when the same update has already
been broadcast by a certain number of neighbours (like for example in
the PSFQ protocol \cite{Wan:Campbell:Krishnamurthy:05}). A further
optimization is to adapt the beaconing rate depending on the
neighbourhood density and the observed rate at which variable updates
arrive, with the goal to maintain a balance between delay, reliability
and medium utilization. Aside from such operational matters, the
functionality of VarDis can be extended for example by allowing more
than one writer for a variable, liveness detection and failure
takeover for variables, or variables with limited scope which are only
disseminated in a $k$-hop neighbourhood.  Finally, one of the next
steps is the implementation of DCP and VarDis on real hardware and
their experimental evaluation. A key part of this implementation will
be the efficient choice of data formats and reduction of packet
overheads.



\end{document}